\documentclass[12pt,preprint]{aastex}
\shorttitle{Local Leo Cold Cloud}
\shortauthors{Meyer et al.\ 2011}
\begin{document}


\title{The Remarkable High Pressure of the Local Leo Cold Cloud}


\author{David M. Meyer\altaffilmark{1,5},
        J. T. Lauroesch\altaffilmark{2},
        J. E. G. Peek\altaffilmark{3,6}, and
        Carl Heiles\altaffilmark{4}}


\altaffiltext{1}{Center for Interdisciplinary Exploration and Research
in Astrophysics, Department of Physics and Astronomy, Northwestern
University, Evanston, IL  60208; davemeyer@northwestern.edu}

\altaffiltext{2}{Department of Physics and Astronomy, University of
Louisville, Louisville, KY  40292}

\altaffiltext{3}{Department of Astronomy, Columbia University, Pupin
Physics Laboratories, 550 West 120th Street, New York, NY  10027}

\altaffiltext{4}{Department of Astronomy, University of California,
Berkeley, 601 Campbell Hall, Berkeley, CA  94720}

\altaffiltext{5}{Visiting Astronomer, Kitt Peak National Observatory,
National Optical Astronomy Observatory, which is operated by the
Association of Universities for Research in Astronomy, Inc. under
cooperative agreement with the National Science Foundation}

\altaffiltext{6}{Hubble Fellow}


\begin{abstract}

Using the Space Telescope Imaging Spectrograph (STIS) onboard the
{\it Hubble Space Telescope}, we have obtained high-resolution
ultraviolet spectra of the C~I absorption toward two stars behind
the Local Leo Cold Cloud (LLCC).  At a distance ($\approx$20~pc)
that places it well inside the Local Bubble, the LLCC is the
nearest example of the coldest known (T~$\approx$~20~K) diffuse
interstellar clouds.  The STIS measurements of the C~I
fine-structure excitation toward HD~85259 and HD~83023 indicate that
the thermal gas pressure of the LLCC is much greater than that
of the warm clouds in the Local Bubble.  The mean LLCC pressure measured
toward these two stars (60,000~cm$^{-3}$~K) implies an H~I density of
$\approx$3000~cm$^{-3}$ and a cloud thickness of $\approx$200~AU at
the 20~K cloud temperature.  Such a thin, cold, dense structure could
arise at the collision interface between converging flows of
warm gas.  However, the measured LLCC pressure is appreciably higher
than that expected in the colliding cloud interpretation given the
velocity and column density constraints on warm clouds
in the HD~85259 and HD~83023 sightlines.
Additional STIS measurements of the Zn~II, Ni~II, and Cr~II
column densities toward HD~85259 indicate that the
LLCC has a modest ``warm cloud'' dust depletion pattern
consistent with its low dust-to-gas ratio determined from
H~I 21~cm and 100 $\micron$ observations.  In support of the
inferred sheet-like geometry for the LLCC, a multi-epoch comparison
of the Na~I absorption toward a high-proper-motion background star
reveals a 40$\%$ column density variation indicative of LLCC Na~I
structure on a scale of $\approx$50~AU.

\end{abstract}





\section{Introduction}

The Local Leo Cold Cloud (LLCC) \citep{peek11a} is the predominant
central component (22 square degrees centered near $l=223\arcdeg$,
$b=+44\arcdeg$) of a narrow ($\sim$2$\arcdeg$), broken ribbon of diffuse
cold gas stretching over 70$\arcdeg$ of the sky at high Galactic latitude
\citep{hau10}.  The LLCC was discovered by \citet{ver69} through its very
narrow H~I 21~cm emission indicative of a kinetic temperature below 30~K.
Utilizing 21 cm observations of three background extragalactic radio
sources, \citet{hei03} revisited the LLCC as part of their Arecibo
Millenium H~I survey and measured spin temperatures of $\approx$20~K
and column densities of $\approx$3~$\times10^{19}$~cm$^{-2}$ for the
intervening gas.  The combination of such temperatures and column
densities are remarkable in that they imply an extremely thin ($<0.1$~pc),
sheetlike geometry for the cloud if it is in thermal pressure
equilibrium.

In the first optical absorption line study of the LLCC, \citet{mey06}
obtained high-resolution observations of interstellar Na~I toward
33 stars in the region and were able to place a firm upper limit of
45~pc on the cloud distance.  This distance constraint places the
LLCC well inside the Local Bubble of low-density gas
surrounding the Sun out to distances of
$\sim$100~pc or more \citep{cox87}.  Based on measurements of the
soft X-ray background \citep{fri80}, it has long been thought that the
Local Bubble is filled with a highly-ionized, hot ($\sim$10$^6$~K)
plasma.  Recently, \citet{peek11a} tested this hypothesis through a
multi-wavelength study utilizing the LLCC in an X-ray shadowing
experiment.  Key steps included tightening the LLCC distance to
between 11.3$\pm$0.2 and 24.3$\pm$0.2~pc based on additional optical
absorption-line measurements and mapping the LLCC H I 21 cm emission
at spatial and velocity resolutions of 4$\arcmin$ and
0.184~km~s$^{-1}$ through the Galactic Arecibo L-Band Feed Array
H~I (GALFA-HI) survey \citep{peek11b}.  In comparing this map (with
a peak H~I column density of $2.5\times10^{20}$~cm$^{-2}$) to the
{\it ROSAT} soft (1/4 keV) X-ray data \citep{sno97} in the LLCC
region, \citet{peek11a} find appreciably less X-ray shadowing than
expected from a standard Local Bubble model.
Based on this result and other
evidence \citep{wel09}, they argue that the isotropic component of
the observed soft X-ray background is mostly a foreground effect due
to solar wind charge exchange with neutral interstellar matter at
the $\sim$100~AU distant heliopause \citep{sto05,kou09}. 

Given the LLCC's location deep inside the Local Bubble, a measure of
its thermal gas pressure would provide a key physical diagnostic of
this cold cloud and its environment.  The warm, partially-ionized
clouds that are known to inhabit the Local Bubble collectively have
an average temperature of 6700~K and a mean thermal pressure (P/k)
of 2300~cm$^{-3}$~K \citep{red04}.  If the LLCC is in pressure
equilibrium with these warm clouds and their environment, one would
expect it to have a similar mean pressure.  However, \citet{red08}
have suggested that this cold cloud is the result of an impact
between warm clouds in the Local Bubble and subsequent compression
at the cloud interface.  Models of the transonic compression of
colliding flows of warm gas have yielded
cold, sheet-like structures at the collision interface
characterized by significant overpressures with respect to the
surrounding medium \citep{aud05,hei06,vaz06}.
Thus, if the LLCC is the product of such a collision,
its pressure should clearly exceed the mean pressure of the warm
clouds in the Local Bubble.

Ultraviolet spectroscopy of the C~I absorption produced by the LLCC
in its background stars has the potential to provide an accurate
measure of the cloud's thermal gas pressure.  Since the
fine-structure states of interstellar C~I are primarily populated
by collisions with hydrogen atoms in diffuse clouds, their excitation
is sensitive to the local gas pressure \citep{jen79}.  The many
C~I transitions observable in the far-UV (key multiplets centered
near 1277, 1280, and 1329~\AA) include lines with a large range
in oscillator strength arising from each of the J~$=$~0,~1,~2
fine-structure levels.  A number of other diagnostic transitions
through which the physical character of the LLCC can be evaluated
are also only observable in the ultraviolet.  However, due in part
to a lack of bright UV background sources, there has been no previous
UV absorption-line study of this cloud.

In this paper, we present high-resolution UV spectra of the LLCC C~I
absorption in two sightlines observed with the Space Telescope
Imaging Spectrograph (STIS) onboard the {\it Hubble Space Telescope}
({\it HST}).  These observations conclusively show that the thermal
gas pressure of the LLCC is well above that of the warm clouds in
the Local Bubble.  Indeed, the measured LLCC pressure is higher than
that expected in the colliding cloud interpretation of the LLCC as a
sheet-like structure at the collision interface.  Nevertheless, in
support of this interpretation, we find that a multi-epoch
comparison of the Na I absorption toward a high-proper-motion
background star reveals significant variation indicative of
LLCC structure on a scale of $\approx$50~AU.

\section{Observations}

Among the 16 stars found by \citet{mey06} to exhibit narrow Na~I
absorption corresponding to the LLCC, three (HD~83023, HD~85259,
and HD~84722) were targeted for {\it HST} STIS observations of
their UV interstellar absorption lines.  These specific stars were
selected because they are the brightest UV background sources
with appreciable LLCC absorption and they are all rapid rotators
($v$sin$i$~$>$~100~km~s$^{-1}$) well-suited for interstellar
absorption-line measurements.  HD~83023 (d~$=$~209$^{+24}_{-20}$~pc)
\citep{van07} and HD~85259 (239$^{+33}_{-25}$~pc) have sky positions
behind the periphery of the LLCC while HD~84722
(84.8$^{+3.8}_{-3.4}$~pc) is located behind the cloud interior
about 52$\arcmin$ away from HD~85259 (which corresponds to a
sightline separation of 0.3~pc at the cloud distance of
$\approx$20~pc).

The STIS observations of HD~83023, HD~84722, and HD~85259 were
obtained in 2010 April, November, and December, respectively.
HD~83023 and HD~85259 were both observed using the E140H and E230H
echelle gratings to cover the 1242~-~1444~\AA\ and 1824~-~2101~\AA\
wavelength regions at a velocity resolution of 2.75~km~s$^{-1}$.
Due to the faintness of HD~84722 in the far-UV, this star was
only observed with the E230H setup.  The observed wavelength
regions were chosen to cover key interstellar transitions of
C~I, C~I*, C~I**, C~II*, S~I, S~II, Cl~I, and Ni II with the
E140H setup and Mg~I, Cr~II, and Zn~II with the E230H setup.
The raw two-dimensional STIS echelle spectra were flat-fielded,
wavelength-calibrated, and scattered-light-corrected using the
standard STSDAS data reduction software.  With net exposures
ranging from 1750~s (E230H) for HD~84722 to 4650~s (E140H) for
both HD~83023 and HD~85259, the final extracted spectra are
characterized by continuum S/N ratios ranging from $\approx$15
at 2050~\AA\ for HD~84722 and at 1260~\AA\ for HD~83023 to
$\approx$80 at 1315~\AA\ for HD~85259.  All of the STIS
observations for this program (GO 11736) were obtained
in a total of 7~{\it HST} orbits.

In support of our STIS data analysis, we re-observed the
interstellar Na~I D$_2$ $\lambda$5889.951 and D$_1$
$\lambda$5895.924 absorption toward HD~83023, HD~84722, and
HD~85259 using the same 0.9~m coude feed telescope and
spectrograph setup as the \citet{mey06} observations at Kitt
Peak National Observatory (KPNO).  These observations were
obtained with the purpose of determining the absorption-line
profile of the LLCC as accurately as possible at higher S/N and
velocity resolution than the STIS observations.  The new exposures
were reduced in the same manner as the earlier data and summed
with these data to achieve final spectra with significantly enhanced
S/N ratios at a measured velocity resolution of 1.3~km~s$^{-1}$.
Specifically, HD~83023 was observed for an additional 7~hr in
2009 January to produce a net Na~I spectrum (10~hr total exposure)
characterized by a continuum S/N ratio of 320.  The high S/N
ratio (340) of the final HD~84722 spectrum (14.75~hr total
exposure) was achieved through new exposures spanning
12.75~hr in 2007 January and 2008 March.  HD~85259 was observed
for an additional 9~hr in 2008 March and 2011 February to
produce a net Na~I spectrum (10~hr total exposure) characterized
by a S/N ratio of 250.

In 2009 January, we also conducted new KPNO Na~I observations of
the star HD~84937 to compare with the 2006 January measurements of
\citet{mey06} and probe the small-scale Na~I structure of the
foreground LLCC\@.  This star is located behind the outer periphery
of the LLCC at a distance of 72.7$^{+4.4}_{-3.9}$~pc and it is
distinguished by a very large proper motion of
0.860~arcsec~yr$^{-1}$ \citep{van07}.  The new exposures of
HD~84937 (9~hr net integration) were reduced in the same manner
as the earlier data (4.5~hr net integration) and the resultant
summed spectra both have S/N ratios of 100 at a velocity
resolution of 1.3~km~s$^{-1}$.  HD~84937 was not
selected as a UV interstellar absorption target for our STIS
observations due to its late spectral type (F5) and faintness
($V=8.28$).

\section{Results}

\subsection{Na I}

In modeling their multi-star observations of the Na~I absorption-line
profiles corresponding to the LLCC, \citet{mey06} found that a
single-component solution provided an excellent fit to the data.
Our significantly higher S/N observations of the LLCC Na~I D$_2$
and D$_1$ profiles toward HD~83023, HD~85259, and HD~84722 shown
in Figures~1, 2, and 3 indicate that a double-component fit is
more appropriate.  The fits to these profiles were made with the
FITS6P iterative Voigt-profile fitting program \citep{wel94} and
took into account the hyperfine splitting of the Na~I~D lines
($\Delta$v~$=$~1.05~km~s$^{-1}$).  Table~1 lists the resulting
Na~I column densities, line widths (b-values), and LSR velocities
for both the single- and double-component models of the measured
line profiles.  The double-component fits are characterized by a
very narrow dominant core superposed on a much broader,
shallower component with a central velocity offset of no more than
0.4~km~s$^{-1}$.  Due to its high S/N ratio, moderate Na~I line
strength, and modest component separation, the HD~83023
spectrum provides the most accurate measure of the line width
(b~$=$~0.24$\pm$0.01~km~s$^{-1}$) of the narrow Na~I component.
This b-value was adopted for the stronger narrow Na~I components
toward HD~85259 and HD~84722 in optimizing the double-component
fits to their Na~I profiles.

In the case of the broader secondary
Na~I component, there are also hints of this feature in other
LLCC sightlines observed by \citet{mey06}, but nothing
definitive due to the lower S/N ratio of their data.
However, as shown in Figure~4, our re-observation of HD~84937
yields a net spectrum of sufficiently enhanced S/N (140) to
provide further evidence of this secondary Na~I component.
The existence of a weak broader component at nearly the same
velocity as that of the narrow LLCC Na~I absorption toward multiple
background targets suggests that the clouds responsible for these
features may be related.  Furthermore, we detect this broader
component in the spectra of HD~84722 (84.8~pc) and HD~84937
(72.7~pc) which are both located inside the Local Bubble.
Among the LLCC background stars observed by
\citet{mey06}, none within 157~pc exhibits the additional Na~I
components seen (at LSR velocities ranging from -11.5 to
-1.5~km~s$^{-1}$) toward some of their more distant stars.
Meyer et al. interpret this additional Na~I absorption as arising
from the neutral gas boundary of the Local Bubble at a distance
between 100 and 150~pc in the direction of the LLCC\@.

Interestingly, \citet{peek11a} also find that a double-component
solution provides the best fit to their H~I 21~cm emission-line
observations of the LLCC\@.  In Table~1, we list the H~I column
densities, line widths (b-values), and LSR velocities derived
from their optically-thin, two-component fits to the LLCC H~I
emission observed at 4$\arcmin$ resolution toward HD~83023,
HD~85259, and HD~84722 (no LLCC H~I emission was detected
toward HD~84937).  In comparing the Na~I and H~I measurements
in these sightlines, it is important to note that the
4~arcmin H~I emission-line beam is much larger than the
10$^{-4}$~arcsec Na~I absorption-line beam.  Also, as a trace
neutral, Na~I (ionization potential of 5.14~eV) is not the
dominant ion of Na in H~I clouds.  Consequently, the column density
of Na~I is a function of both the H~I column and environmental
parameters such as the cloud temperature, electron density, and
radiation field governing the Na ionization equilibrium.
Nevertheless, as compared to Na~I observations of other Galactic
H~I clouds \citep{wel07}, the stronger LLCC components toward
HD~83023, HD~85259, and HD~84722 all have among the highest Na~I
columns measured in such low N(H~I) sightlines.  Since the
recombination of trace neutrals scales with cloud temperature as
$T^{-1.6}$ for a given N(H~I) and constant pressure \citep{hei07},
such high Na~I columns would be expected in an H~I cloud as cold
as the LLCC\@.  Assuming that the turbulent Na~I and H~I motions
in the stronger LLCC components are similar, a comparison of the
Na~I (0.24~km~s$^{-1}$) and H~I (mean value of 0.59~km~s$^{-1}$)
linewidths through the expression $b^2=(2kT/m)+2v_t^2$
yields a cloud kinetic temperature of 18~K and a one-dimensional
rms turbulent velocity ($v_t$) of 0.16~km~s$^{-1}$.

The principal differences between the LLCC Na~I and H~I profiles
lie in the weaker of the two fitted components.  In all three
sightlines, the weaker Na~I component is much broader than the
weaker H~I component.  Furthermore, the velocity separations
between the stronger and weaker components are greater for H~I than
Na~I and the column density differences between the components
are more pronounced in the Na~I fits than those for H~I.  The
most likely explanation for these differences are that the weaker
Na~I and H~I components are not sampling the same parcels of gas.
In the case of the weaker Na~I component, its low column and
broad width would make the corresponding H~I emission difficult
to detect in the \citet{peek11a} data.  Such a weak H~I feature
would have a FWHM of $\approx$13~km~s$^{-1}$ and blend in with the
broad red wing of the H~I emission associated with the neutral
gas boundary of the Local Bubble.  \citet{peek11a} interpret the
similarly narrow H~I components in their two-component fits to
the LLCC H~I profile as arising from two colliding clouds.  In
this picture, the absence of Na~I absorption corresponding to
the lower H~I column cloud could reflect an environmental factor
such as a lower electron density.  Alternatively, the narrow
Na~I absorption could arise from high-density gas at the
collision interface between the two clouds.  In either case, 
the broad Na~I component could arise from warmer diffuse gas
associated with the two H~I clouds or the post-shock material
just outside the cold interface layer.

\subsection{C I}

The STIS observations of the LLCC C~I absorption toward
HD~85259 and HD~83023 support the two-component solution to
the optical Na~I profiles.  As illustrated in Figures~5 and 6,
spectra of the 1329~\AA\ multiplet reveal significant C~I
absorption arising from each of the J = 0, 1, 2 fine-structure
states toward both HD~85259 and HD~83023.  In the case of
HD~85259, the C~I lines are strong enough and the spectral
S/N ratio is high enough to fully utilize the C~I multiplets
centered near 1277 and 1280~\AA\ along with the 1329 \AA\ data
in measuring the C~I, C~I*, and C~I** profiles and column
densities.  The best simultaneous FITS6P fit to these data
yields a two-component solution for the C~I profile characterized
by a dominant narrow (b~$=$~0.4$\pm$0.1~km~s$^{-1}$) component
and a weaker broad (b~$=$~2.6$\pm$0.4~km~s$^{-1}$) component
with the same velocity separation as the Na~I profile fit.
The C~I* and C~I** profiles were fit by a single component
with the same b-value as the narrow C~I component.  In the case
of HD~83023, the low S/N ratio ($<$25) of its STIS spectra
shortward of 1300~\AA\ makes the C~I 1277 and 1280~\AA\ multiplets
of limited utility in fitting the C~I, C~I*, and C~I** profiles.
Nevertheless, we find that a two-component solution consisting
of a narrow (b~$=$~0.4~km~s$^{-1}$) component and a weaker
broad (b~$=$~3.1$\pm$0.6~km~s$^{-1}$) component with the same
velocity separation as the Na~I fit provides an excellent fit
to the LLCC C~I absorption toward HD~83023.  As in the case of
HD~85259, the C~I* and C~I** profiles toward HD~83023 are also
best fit by a single narrow component with a b-value of
0.4~km~s$^{-1}$.  Figures~5 and 6 display all of these fits
along with the observed profiles near 1329~\AA\@.  The resultant
C~I, C~I*, and C~I** column densities for both the narrow and
broad LLCC components toward HD~85259 and HD~83023 are listed
in Table~2.  These column densities were calculated using the
updated C~I oscillator strengths determined by \citet{jen11,jen01}.

\subsection{Other Neutrals and Ions}

In addition to C~I, the STIS spectra reveal weak LLCC absorption
arising from other trace neutrals such as S~I, Mg~I, and Cl~I.
Figures~7 and 8 compare the S~I $\lambda$1425.030, Mg~I
$\lambda$2026.477, and Cl~I $\lambda$1347.240 profiles to those
of C~I and Na~I toward HD~85259 and HD~83023.  The C~I
$\lambda$1276.483 and $\lambda$1280.135 profiles illustrated in
Figures~7 and 8 are respectively the weakest C~I lines detected
in the HD~85259 and HD~83023 spectra and are fit with the same
two-component solutions shown for the strong C~I
$\lambda$1328.833 line in Figures~5 and 6.  There is no evidence
of a broad component in the weak S~I, Mg~I, and Cl~I absorption
toward either star.  These lines were all fit using a single
component with the same b-value (0.4~km~s$^{-1}$) as the narrow C~I
component.  The resultant S~I, Mg~I, and Cl~I column densities
are listed in Table~2.  Given the weakness of the lines, these
column densities are relatively insensitive to the profile fit.
They were derived utilizing the S~I, Mg~I, and Cl~I oscillator
strengths compiled by \citet{mor03}.

Figures~9 and 10 compare the interstellar absorption profile of
Cl~I with those of the dominant ions S~II, Cr~II, Ni~II, and
Zn~II plus that of the excited fine-structure fraction of
C~II (C~II*) toward HD~85259 and HD~83023.  The C~II* and dominant
ion profiles clearly exhibit multiple velocity components
at LSR velocities blueward of the narrow LLCC absorption
benchmarked by Cl~I\@.  These blue components are especially
strong toward HD~85259 and are detectable in its Na~I spectrum
(see Figure~7).  As discussed earlier, these components presumably
arise from clouds associated with
the neutral gas boundary of the Local Bubble through
which HD~85259 and HD~83023 are distant enough to sample in
absorption.  Although the STIS spectrum of the nearer star HD~84722
is more limited in wavelength coverage and S/N ratio, it does tend to
support this interpretation.  Specifically, as shown in Figure~11,
the Zn~II $\lambda\lambda$2026.136, 2062.664 doublet observed
toward HD~84722 does not exhibit the strong components seen
blueward of the LLCC absorption toward HD~85259 and HD~83023.  At
the same time, there is a hint of Zn~II absorption corresponding to a
very weak Na~I component seen at $v_{LSR}$~$=$~-8.1~km~s$^{-1}$
in the high S/N optical spectrum of HD~84722.  Whether this
component is associated with the LLCC, the HD~84722 environment,
or somewhere in between, it most likely
arises from gas inside the Local Bubble.
In general, due to their velocity separation, the strong
non-LLCC components are typically not a major complication in
fitting the LLCC portion of the dominant ion profiles
toward HD~85259 and HD~83023.  The
LLCC S~II, Cr~II, Ni~II, Zn~II, and C~II* column densities
derived from these fits (using f-values from \citet{jen06}
for Ni~II and from \citet{mor03} for S~II, Cr~II, Zn~II,
and C~II*) are tabulated in Table~2.

Among the dominant ion lines observed toward HD~85259 and HD~83023,
the Zn~II doublet provides the best profile model of the LLCC
absorption due to its modest strength and saturation.  In the
case of HD~85259, its LLCC Zn~II absorption is well fit by a
single narrow component with a b-value of 0.4$\pm$0.1~km~s$^{-1}$.
In contrast, the weaker LLCC Zn~II absorption toward HD~83023
is best fit with a two-component model consisting of a narrow
(b~$=$~0.4~km~s$^{-1}$) core and a shallower broad
(b~$=$~2.8~km~s$^{-1}$) component with the same velocity
separation as its Na~I and C~I fits.  Fitting the much stronger
S~II $\lambda\lambda$1250.584, 1253.811, 1259.519 triplet is
more problematic, particularly for HD~83023 due to the low S/N
ratio of its far-UV spectrum.  Nevertheless, we find that the
best fits to the LLCC S~II absorption toward HD~85259 and HD~83023
both require a narrow and a broad component.  Due to the saturation
of the narrow component, these fits are relatively insensitive
to its column density.  Consequently, we adopted a b-value of
0.4~km~s$^{-1}$ for this S~II component toward both stars and 
fixed its column density based on the measured Zn~II
columns, the solar S/Zn elemental
abundance ratio (log(S/Zn)~$=$~2.56) \citep{lod03}, and a Zn
depletion into dust of 0.1~dex relative to S \citep{wel99,jen09a}.
The resulting S~II column densities listed in Table~2 for the
broad LLCC component were determined using fits with the same
component separations as the Na~I and C~I models.  In the case
of the strong C~II* $\lambda$1335.708 (f~$=$~0.115)
absorption toward HD~85859 and HD~83023,
a double-component LLCC solution with a narrow
saturated core is also warranted.  Unfortunately, the adjacent
C~II* $\lambda$1335.663 (f~$=$~0.0128) line
does not provide much of a fitting
constraint on the column density of this narrow component due
to its blend with the strong non-LLCC $\lambda$1335.708
components.  Thus, the resulting column densities listed in
Table~2 for the narrow LLCC C~II* components toward HD~85259
and HD~83023 are very uncertain.

\section{Discussion}

\subsection{The LLCC C~I Fine-Structure Excitation}

With excitation energies ($E/k$) of 23.6 and 62.4~K, the J~$=$~1
and J~$=$~2 fine-structure states of C~I are predominantly
populated through collisions with hydrogen atoms in cold
interstellar H~I clouds.  Since the collision rates are a function
of the H~I density and temperature, the measured C~I excitation
reflects the thermal gas pressure of these clouds.
The most thorough analysis to date of the C~I fine-structure
excitation in the Galactic interstellar medium has been carried
out by \citet{jen11} who show that this excitation quickly
equilibrates to its environmental pressure on a timescale
of less than one year.  Utilizing high-resolution {\it HST}
STIS observations of the far-UV C~I multiplets, they measure
the C~I, C~I*, and C~I** column densities as a function of
velocity in the interstellar absorption-line profiles of
89~stars.  Building upon the smaller sample of their earlier work
\citep{jen01}, \citet{jen11} calculate the thermal
pressures of 2416 parcels of interstellar gas based on these
column densities.  Following the approach of \citet{jen79},
they utilize the quantities $f1=N$(C~I*)/$N$(C~I$_{total}$)
and $f2=N$(C~I**)/$N$(C~I$_{total}$) to relate the measured
C~I excitation to a grid of derived gas pressures as a function
of cloud temperature.  They find that the excitation of their
large sample of clouds averages out to $f1=$0.209 and
$f2=$0.068 with a derived mean gas pressure ($p/k$) of
3800~cm$^{-3}$~K\@.

Our measurements of the LLCC C~I, C~I*, and C~I** column
densities toward HD~85259 and HD~83023 yield respective
($f1$,~$f2$) values of (0.396$\pm$0.036, 0.069$\pm$0.008) and
(0.366$\pm$0.041, 0.100$\pm$0.018).  In Figure~12, we plot
these points plus their weighted mean value
(0.383$\pm$0.027, 0.074$\pm$0.007)
amidst three curves denoting the derived
($f1$,~$f2$) values for a range of gas pressures in clouds
with temperatures of 15, 20, and 30~K (as cloud temperatures
rise above these values, the curves remain closely spaced at
low pressure (low~$f1$) and pull back to the left
at high pressure (high~$f2$)).  These curves were
calculated in a manner similar to that of \citet{jen11}
using the same collision, radiative decay, and optical pumping
rates along with the average stellar radiation field
appropriate for the solar vicinity \citep{mat83}.  As noted
by \citet{jen01}, the ($f1$,~$f2$) values are relatively
insensitive to uncertainties in the radiation field when
the pressure is greater than 10,000~cm$^{-3}$~K\@.
Given that the LLCC has a temperature of
$\approx$20~K, Figure~12 clearly indicates that its
gas pressure toward both HD~85259 and HD~83023 is well in
excess of 10,000~cm$^{-3}$~K\@.  Specifically, the HD~83023 data
point falls 1.6$\sigma$ above the 20~K curve at a pressure of
40,000~cm$^{-3}$~K and that of HD~85259 falls 1.8$\sigma$ below
the 20~K curve at a pressure of 80,000~cm$^{-3}$~K\@.  Their
weighted mean ($f1$,~$f2$) value falls almost exactly on the
20~K curve at a pressure of 60,000~cm$^{-3}$~K\@.
In contrast, the vast majority of the
\citet{jen11} data points have ($f1$,~$f2$) values that place
them above and/or left of the theoretical curves for individual
clouds.  They interpret these values as arising from mixtures
of low-pressure and high-pressure C~I components in their
long-path sightlines that unresolvably overlap in velocity.
In the cases of HD~83023 and HD~85259, the
narrow LLCC C~I component stands out in the spectra of these
nearby stars with an excitation that closely matches the
theoretical expectations for a very cold, highly-pressurized
cloud.  Such conditions are very rare among the 2416 C~I
parcels sampled by \citet{jen11} - less than 1$\%$ of these
clouds have ($f1$,~$f2$) excitation values corresponding to
temperatures less than 30~K and pressures greater than
10,000~cm$^{-3}$~K\@.

The highly-pressurized nature of the LLCC is particularly
intriguing given its location deep inside the Local Bubble.
As reviewed by \citet{jen09b} and \citet{wel09}, the pressure
balance of interstellar gas in this environment has
long been a contentious issue.  Since the discovery of the
diffuse soft X-ray background in the 1970s, numerous studies
through the 1990s interpreted it in terms of hot gas filling
the Local Bubble with a pressure somewhere between 10,000 and
20,000~cm$^{-3}$~K\@.  Yet, studies of cooler clouds within
the Bubble have consistently yielded pressures below
these values.  \citet{jen02} utilized {\it HST} C~I
observations to constrain the thermal pressures of
four clouds inside or near the edge of the Local Bubble to
values between 1000 and 10,000~cm$^{-3}$~K\@.  \citet{red04}
measured the temperatures of 50 warm, partially-ionized
clouds within the Bubble and determined their mean
thermal pressure to be 2300~cm$^{-3}$~K\@.  Recently,
measurements of the foreground X-ray emission
from solar wind charge exchange \citep{rob10}
and the \citet{peek11a} finding of weak X-ray shadowing
by the LLCC have weakened the case for a higher-pressure hot
Local Bubble.  Although some hot gas is undoubtedly
present, its thermal pressure is most likely in a range
(3000 to 7000~cm$^{-3}$~K) consistent with that
of the warm clouds \citep{fri11}.  In contrast, the much
higher mean LLCC pressure of 60,000~cm$^{-3}$~K measured toward
HD~83023 and HD~85259 clearly indicates that the LLCC is not in
thermal pressure equilibrium with either the hot gas or the warm
clouds in the Local Bubble.  Given the LLCC temperature of
$\approx$20~K and a typical H~I column density of
$\approx$10$^{19}$~cm$^{-2}$, this pressure corresponds to a
cloud with a density of $\approx$3000~cm$^{-3}$ and a thickness
of $\approx$200~AU ($\approx$0.001~pc).  Note that even for such
a thin cloud, the C~I excitation equilibrates to the cloud
pressure on a timescale ($<$~1~yr) that is much shorter than
the LLCC's line-of-sight turbulent crossing time
($\approx$6000~yr).

Our finding that the LLCC is significantly overpressured
with respect to the Local Bubble is
qualitatively consistent with predictions that the LLCC
and its anomalously low temperature could be the result of a
warm cloud collision \citep{vaz06,mey06,red08}.  In their
one-dimensional numerical simulations of the large-scale transonic
(Mach~number~$\sim$~1) compression of colliding warm gas flows,
\citet{vaz06} find that a thin cold layer forms within
the shocked interface.  This layer is characterized by a
temperature of $\sim$25~K, pressure of $\sim$6650~cm$^{-3}$~K,
thickness of $\sim$0.03~pc, and H~column density of
$\sim$$2.5\times10^{19}$~cm$^{-2}$ after $\sim$1~Myr of evolution.
In general, the numerical models of \citet{vaz06} show that the
cold, dense gas at the collision interface is overpressured with
respect to the infalling warm gas by factors of 1.5 to 5.
In contrast, the C~I excitation measured toward HD~83023 and
HD~85259 indicates that the LLCC is overpressured by a factor of
$\approx$30 with respect to the warm clouds in the Local Bubble.
Such a large discrepancy provides a significant challenge toward
quantitatively understanding the LLCC in a colliding cloud context.

A key consideration in achieving an LLCC-type pressure through
a warm cloud collision is the velocity differential between
the colliding clouds since the pressure of the cold interface
layer should balance the total (thermal plus ram) pressure
of the converging flows.  In their analytical model of two
oppositely-directed warm gas streams (with $T~=~7100$~K and
$n~=~0.34$~cm$^{-3}$), \citet{vaz06} find that an inflow Mach
number of 3.7 is required of each stream relative to the
interface layer to produce a pressure of 60,000~cm$^{-3}$~K in
that layer.  Such an inflow would correspond to a velocity
difference of 60~km~s$^{-1}$ between the streams given an
8.4~km~s$^{-1}$ sound speed in the 7100~K gas.  In terms of
radial velocity, there is no evidence of warm clouds in the
LLCC vicinity with such a large velocity differential.  In
Figure~13, we compare the H~I~21~cm emission and the C~II
$\lambda$1334.532 absorption in the HD~85259 and HD~83023
sightlines.  This strong C~II line is a very sensitive tracer
of both warm and cold interstellar gas due to its high $f$-value,
the abundance of C, and the C~II ionization potential (24.38 eV).
The maximum velocity width of the single saturated C~II
$\lambda$1334.532 absorption feature observed toward HD~83023
and HD~85259 is about 35~km~s$^{-1}$.  The absence of any other
C~II absorption lines in these spectra indicates that there
are no foreground warm cloud pairs in the LLCC vicinity with
hydrogen column densities greater than 10$^{16}$~cm$^{-2}$ and
a radial velocity differential greater than 35~km~s$^{-1}$.
As noted by \citet{red08}, the LLCC sky position is near the
boundaries of the nearby Gem, Leo, and Aur warm clouds.  Although
the Leo and Aur clouds both have radial velocities similar to
that of the LLCC, the Gem cloud's radial motion is only faster
by 12~km~s$^{-1}$.  Furthermore, the upper limits on the distances
of the Gem, Leo, and Aur clouds are all below the lower bound
(11.26$\pm$0.21~pc) on the LLCC distance \citep{peek11a}.

Other factors may help to provide a better match of the observed
LLCC pressure with that expected in a colliding flow scenario.
For example, the colliding warm clouds might have a significant
difference in their transverse velocities.  Given the measured
radial velocity gradient along the long cold cloud ribbon centered
on the LLCC \citep{hau10}, \citet{hau12} has calculated a
transverse LLCC LSR velocity of 16~km~s$^{-1}$ based on a ring
model of the cloud ribbon.  Although this value is not a serious
constraint on the transverse velocity differential of any parent
warm cloud collision, it is four times greater than the radial
LLCC LSR velocity.  Other factors to consider are the temperature
and density of the warm clouds themselves.  If the temperature
assumed by \citet{vaz06} is decreased with a corresponding increase
of the density (to maintain the same warm cloud pressure), the
velocity flow differential needed to produce a pressure of
60,000~cm$^{-3}$~K in the cold interface layer is reduced by the
square root of the temperature reduction factor.  Although such
adjustments and transverse component considerations can close
the gap, there is no question that the observed radial velocity
constraints are a key hurdle toward fully understanding the
high LLCC pressure in terms of a warm cloud collision.
It is also important to note that the peak LLCC H~I column
density would require warm source clouds of appreciably higher
column density than any previously identified in the nearby Local
Bubble.  Although the broad absorption component seen in our
Na I and UV spectra attests to an appreciable column of warm gas
in the LLCC sightlines, it could be associated with post-shock
gas just outside the 20~K interface layer rather than the warmer
source clouds.

A common feature of the colliding warm flow models
\citep{aud05,gaz05,hei06,vaz06} is the turbulent fragmentation
of the cold interface layer into clumpy structures.  Some of
the simulations \citep{aud05,gaz05} have shown that localized
regions within these structures can reach pressures up to
10$^5$~cm$^{-3}$~K even when the collision is transonic
\citep{hen07}.  It is quite likely that the different C~I
LLCC pressures measured toward HD~83023 (40,000~cm$^{-3}$~K) and
HD~85259 (80,000~cm$^{-3}$~K) (both assuming $T$~$=$~20~K) are due in
part to real localized pressure variations rather than measurement
errors alone.  Both sightlines are located on the periphery of the
LLCC (4.4$\arcdeg$ (1.4~pc) apart) with H~I column densities that
are more than 10 times below the peak interior value.  Given the
narrowness of their C~I absorption beams and the pressure-implied
thinness of their cloud regions, the HD~83023 and HD~85259
sightlines clearly provide extremely-localized pressure samples
of the LLCC\@.  Consequently, the agreement of their significant
overpressures within a factor of two argues for a large-scale
LLCC explanation like a warm cloud collision.  At the same
time, the turbulent fragmentation expected in the wake of such
a collision could push these localized overpressures above the
limiting values implied by the measured velocity constraints
on the colliding clouds.

\subsection{The Elemental Dust Depletion of the LLCC}

The abundance and elemental composition of dust in diffuse
interstellar clouds can be estimated by comparing the gas-phase
column density ratios of various dominant ions measured from their
UV absorption lines to a set of reference elemental abundance
ratios such as the solar values.  As reviewed by \citet{sav96},
a number of interstellar UV studies have shown that some elements
such as Zn are only weakly depleted from the gas phase into dust
grains while others such as Ni and Cr exhibit strong depletions.
Recently, \citet{jen09a} has systematically analyzed the dust
depletion patterns of 17 elements based on UV absorption-line
measurements toward 243 stars and the \citet{lod03} solar
reference abundances.  He interprets the depletion of
each element relative to hydrogen in terms of a ``line-of-sight
depletion strength factor'' ($F_*$) applicable to all of the
elements in a given sightline plus two other element-specific
parameters.  $F_*$ is a dimensionless parameter that Jenkins has
normalized to equal 0.0 for the lowest collective depletions
in his sample and to equal 1.0 for the strong depletions observed
in the cold $\zeta$~Oph cloud that has long been considered the
prototype for the cold neutral medium.  The \citet{jen09a} results
allow us to analyze the measurements of the LLCC gas-phase
elemental abundances and interpret their dust depletion in the
comparative context of a large sample of other Galactic sightlines.
In our analysis, the cold hydrogen gas comprising the LLCC is
entirely in the form of H~I (we show in $\S$4.4 that the
H$_2$ concentration is very low) and the LLCC column density
ratios of dominant ions are taken to reflect their gas-phase
elemental abundance ratios.

As listed in Table~2, the Zn~II and Ni~II lines observed toward
HD~85259 provide the best-measured dominant ion column densities
from our STIS spectra for the gas associated with the narrow LLCC
absorption component.  Since Ni is depleted much more strongly
into dust than Zn, the logarithmic ratio of their column densities
relative to the solar Ni/Zn abundance ratio yields a relative
depletion [Ni/Zn]~$=$~log(N(Ni~II)/N(Zn~II))~$-$~log(Ni/Zn)$_{\sun}$
that is very sensitive to the depletion strength factor $F_*$.
Although such elemental comparisons can be compromised by
photoionization to higher ionization stages in sightlines with H~I
columns below 10$^{19.5}$~cm$^{-2}$ like HD~85259 and HD~83023
\citep{jen09a}, it is unlikely in the case of Ni II and Zn II since
they have nearly identical ionization potentials.
The corresponding value of -1.18$\pm$0.10 for [Ni/Zn] in the
HD~85259 sightline indicates a factor of $F_*$~$=$~0.2$\pm$0.1
for the LLCC\@.  In the case of HD~83023, the 2$\sigma$ upper
limit on its LLCC Ni~II column density leads to an upper limit on
[Ni/Zn] that is too large to seriously constrain $F_*$.  Since
Cr is also depleted more strongly into dust than Zn, the
[Cr/Zn] ratio can provide an additional indicator of $F_*$.
Unfortunately, our measurements of the Cr~II column density in
the LLCC consist only of a marginal result toward HD~85259
and upper limits toward HD~83023 and HD~84722.  Nevertheless,
the resulting [Cr/Zn] values of -1.47$\pm$0.28, $<$-1.05$\pm$0.15,
and $<$-1.48$\pm$0.16 for the LLCC in the HD~85259, HD~83023,
and HD~84722 sightlines correspond to respective $F_*$ factors
and lower limits of 0.7$\pm$0.3, $>$0.2$\pm$0.2, and
$>$0.7$\pm$0.2.  Collectively, these [Cr/Zn] constraints
indicate a somewhat higher $F_*$ for the LLCC than the more
accurate value provided by [Ni/Zn] toward HD~85259.  Specifically,
the [Ni/Zn] and [Cr/Zn] depletions taken together yield a
weighted mean $F_*$ of 0.3 for the LLCC in the HD~85259
sightline.

The modest dust content implied by a low depletion strength
factor for the LLCC is consistent with the infrared dust emission
measured by \citet{peek11a}.  In comparing the 100~$\mu$m emission
and H~I column density over the sky area of the LLCC, they find
a best-fit value of $4.8\times10^{-21}$~MJy~sr$^{-1}$~cm$^{-2}$
for the LLCC dust-to-gas ratio.  This value is lower than the
standard Galactic ($1.0\times10^{-20}$~MJy~sr$^{-1}$~cm$^{-2}$)
and high-latitude ($6.7\times10^{-21}$~MJy~sr$^{-1}$~cm$^{-2}$)
dust-to-gas ratios measured by \citet{bou88} and \citet{sch98},
respectively.  As indicated by $F_*$, the LLCC [Ni/Zn] and [Cr/Zn]
depletions toward HD~85259 are also below those expected for
cold, dense clouds.  They are more consistent with the
[Ni/Zn]~$\approx$~-1.2 and [Cr/Zn]~$\approx$~-1.0 values
typical of warm clouds \citep{wel99}.  Such a pattern would
support the idea that the LLCC is the product of a warm cloud
collision.  Whether or not the collision itself disrupts some
grain material into the gas phase, the LLCC should not exhibit
higher dust depletions than the warm clouds feeding it.

Since Zn is weakly depleted into grains, the value of [Zn/H] is
relatively insensitive to $F_*$ and can be used with the
measured Zn~II column densities to calculate the corresponding
H~I column densities of the gas associated with the narrow
LLCC component.  Applying the
[Zn/H] depletion (-0.1~dex) appropriate for an $F_*$~$=$~0.3
cloud, the Zn~II columns toward HD~85259, HD~83023, and HD~84722
indicate respective LLCC N(H~I) columns of $8.9\pm0.9\times10^{18}$,
$3.5\pm1.0\times10^{18}$, and $3.7\pm1.1\times10^{19}$~cm$^{-2}$.
Within the errors, these N(H~I) values are consistent with those
measured for the strongest 21~cm component in the
\citet{peek11a} two-component fit (Table~1) to the LLCC H~I
emission in the three sightlines.  Although there is no evidence of
another narrow component in the Zn II profiles corresponding to
the weaker of the two closely-spaced 21~cm components, the STIS
spectra are not definitive in this regard due to their
lower velocity resolution.  However, for the broad secondary
LLCC component observed in Na~I and C~I, the STIS data do provide
well-measured column densities of S~II and Zn~II toward HD~83023
and of S~II toward HD~85259.  The S~II and Zn~II columns are
particularly interesting to compare because both S and Zn are
weakly depleted into grains in diffuse clouds \citep{sav96}.
Assuming that Zn is depleted into dust by 0.1~dex with respect
to S \citep{wel99,jen09a} and that none of the hydrogen
in the broad component cloud is ionized, the value of
log~N(S~II)/N(Zn~II) in the cloud should equal 2.66.  In the
case of the HD~83023 sightline, the measured value of this
logarithmic column density ratio is 3.00$\pm$0.09.  This
enhancement of the S~II abundance relative to that of Zn
could reflect a partial ionization of hydrogen in the broad
component cloud.  Since S~II (23.34~eV) has a higher ionization
potential than Zn~II (17.96~eV), the ionized fraction of the
cloud could raise the net N(S~II)/N(Zn~II) ratio above that
expected from the neutral fraction.  Such a possibility would
be consistent with the idea of the broad LLCC component arising
from warm post-shock gas or one of the colliding warm clouds producing
the LLCC.  Given the N(S~II)/N(Zn~II) ratio observed for the broad
component toward HD~83023, the non-detection of the corresponding
Zn~II component toward HD~85259 is consistent with its lower
S~II column density.

\subsection{The Electron Density of the LLCC}

Given how the LLCC stands out among diffuse clouds in terms of
its temperature and gas pressure, an evaluation of its
electron density provides another key parameter for comparison.
As reviewed by \citet{wel03},
the electron density of a diffuse H~I cloud can be estimated
by comparing the trace neutral and dominant ion column densities of
various elements under the assumption of photoionization equilibrium.
This comparison is quantified through the expression
$n_e$~=~($\Gamma$/$\alpha$)(N(X~I)/N(X~II))
where $\Gamma$ represents the photoionization rate appropriate
for element X and the local radiation field while $\alpha$ is the
temperature-dependent radiative recombination coefficient associated
with the dominant ion.  Various absorption-line studies
utilizing this approach
have found that the electron densities derived from different
elements in the same diffuse cloud sightline can
differ by as much as a factor of ten \citep{fit97,wel99,wel03}.
Also, these derived electron densities are typically higher
than the $n_e$~$\sim$~10$^{-4}n_H$ value expected under the simplest
assumption that the photoionization of carbon is the predominant
source of electrons in the cloud.  Indeed, \citet{wel03} have
found no significant relationship between $n_e$ (as determined
from the photoionization equilibrium of C, Na, and K) and $n_H$
(as determined from the C~I fine-structure excitation) in their
sample of diffuse cloud sightlines.  Apart from uncertainties in
the input parameters (rate coefficients, radiation field, etc.),
they concluded that
``...additional processes besides photoionization and radiative
recombination commonly and significantly affect the ionization
balance of heavy elements in diffuse interstellar clouds''.
Such processes include charge exchange between the dominant ions
and dust grains (or PAHs) in the cloud
which would raise the trace neutral
abundances and the inferred electron densities \citep{lep88}.

In evaluating the electron density of the LLCC, our focus is
on the HD~85259 sightline due to the greater accuracy of its
narrow-component column density measurements.  We consider first
the photoionization balance of C, S, Na, and Mg.  Although we have
measured only the trace neutral column densities for these species
(Table~2), the corresponding dominant ion columns can be estimated
utilizing the measured Zn~II column, the \citet{lod03} solar
abundances, and the appropriate dust depletions.  In the case of
the latter, our evaluation of the LLCC dust depletion toward
HD~85259 indicates that Zn, C, Na, and Mg are depleted
by 0.10, 0.14, 0.40, and 0.57~dex with respect to S (which is
assumed to be undepleted) \citep{wel99,jen09a}.  The resulting
gas-phase C~II, S~II, Na~II, and Mg~II column densities are
1.80$\times$10$^{15}$, 1.62$\times$10$^{14}$, 7.75$\times$10$^{12}$,
and 9.93$\times$10$^{13}$~cm$^{-2}$, respectively.  The radiative
recombination rate coefficients ($\alpha$) appropriate for
C~II (2.27$\times$10$^{-11}$~cm$^3$~s$^{-1}$),
S~II (2.06$\times$10$^{-11}$~cm$^3$~s$^{-1}$),
Na~II (1.83$\times$10$^{-11}$~cm$^3$~s$^{-1}$), and
Mg~II (2.84$\times$10$^{-11}$~cm$^3$~s$^{-1}$) at the LLCC
temperature of 20~K are taken from the compilation of
\citet{ver99}.  The photoionization rates ($\gamma$) adopted for
C~I (2.58$\times$10$^{-10}$~s$^{-1}$),
S~I (9.25$\times$10$^{-10}$~s$^{-1}$),
Na~I (7.59$\times$10$^{-12}$~s$^{-1}$), and
Mg~I (5.39$\times$10$^{-11}$~s$^{-1}$) were calculated by
\citet{dra11} utilizing the \citet{mat83} estimate of the local
interstellar radiation field.  Assuming photoionization equilibrium
governed by these rates and coefficients, the measured trace
neutral and inferred dominant ion column densities of C, S, Na,
and Mg in the HD~85259 sightline indicate respective LLCC electron
densities of 0.15, 0.20, 0.030, and 0.014~cm$^{-3}$.

Although the spread in the $n_e$ values toward HD~85259 is similar
to those derived from the photoionization balance of C, S, Na, and
Mg in other diffuse cloud sightlines \citep{wel03}, there is an
unusual pattern in the LLCC numbers.  Specifically, the species
(C~I and S~I) with the highest photoionization rates
and ionization potentials yield significantly larger electron
densities than the the two (Na~I and Mg~I) with the lowest rates and
potentials.  Possible explanations for such a pattern range from
a softer-than-assumed local radiation field to biased trace neutral
formation associated with a sharp density peak within the cloud.
As discussed by \citet{lau03}, the latter possibility could produce
ionization-dependent spatial variations in the distribution of
trace neutrals due to the distance differences that various neutrals
could travel on average from their predominant density-peak origin
before photoionization.  Consequently, in the presence of a sharp
density peak, the trace neutrals with the highest photoionization
rates will be more localized near the peak than those with lower
rates and thus, their photoionization balance with density-dependent
recombination will yield higher inferred electron densities.  The
potential attraction of this idea in the case of the LLCC is that
such a density peak would be expected in the colliding warm cloud
model.

The mean LLCC electron density determined from the photoionization
balance of C, S, Na, and Mg toward HD~85259 is 0.1~cm$^{-3}$.
In comparison, an electron density of $\sim$0.4~cm$^{-3}$ would be
expected based on the LLCC hydrogen density
($n_H$~$\approx$~4000~cm$^{-3}$) determined from the C~I
fine-structure excitation in this sightline and the assumption that
the photoionization of carbon is the predominant source of
electrons ($n_e$~$\sim$~10$^{-4}n_H$) in this cloud.  The contrast
between the LLCC and similarly-measured diffuse clouds
is especially striking in terms of $n_e$/$n_H$.
The LLCC value of 2.5$\times$10$^{-5}$ for
this quantity is well outside the $n_e$/$n_H$~$\sim$~0.0002-0.02
range typical of other clouds \citep{wel03}.  Given the modest
dust content of the LLCC, it is certainly possible that the LLCC
lacks the grains (and PAHs) to elevate its trace neutral abundances
and inferred electron density through charge exchange with
dominant ions.  It is more likely that the low LLCC $n_e$/$n_H$
value is due to density inhomogeneities in the cloud such as a
sharp density peak.  The value of $n_H$ is determined from the
C~I fine-structure excitation and, thus, is biased toward high
density regions along the sightline where C~I is most abundant.
The value of $n_e$, on the other hand, is weighted by the broader
sightline distribution of the dominant ions and represents more of
an average throughout the cloud.  Consequently, one would expect
a lower value of $n_e$/$n_H$ in a cloud with a sharp density peak
than one that is more homogeneous.

Another approach to estimating the electron density of a diffuse
cloud is through the analysis of its C~II fine-structure excitation
\citep{fit97,wel99,wel07,dra11}.  At the temperature
(T~$\approx$~80~K) and density ($n_H$~$\approx$~10~cm$^{-3}$)
typical of cold diffuse clouds, this excitation is dominated by
collisions with electrons.  However, in a high-density cloud
like the LLCC, the contribution from collisions with hydrogen
atoms is also significant.  In evaluating the LLCC C~II
fine-structure excitation in the HD~85259 sightline, the largest
source of error is the C~II$^*$ column density.  As discussed in
$\S$3.3, the profile fit to the narrow
C~II$^*$~$\lambda\lambda$1335.663,1335.708 absorption arising from
the LLCC is complicated by heavy saturation and blending issues.
Our best estimate of 3$\times$10$^{13}$~cm$^{-2}$ for N(C~II$^*$)
in this LLCC sightline is thus quite uncertain.  Nevertheless,
utilizing the appropriate excitation rates \citep{wel99} for a
cloud temperature of 20~K and assuming that $n_H$~$=$~4000~cm$^{-3}$,
this level of C~II fine-structure excitation
(N(C~II$^*$)/N(C~II)~$=$~0.0167) indicates an electron density
of 8.8~cm$^{-3}$.
A much smaller value ($n_e$~$=$~0.1~cm$^{-3}$) consistent with
the photoionization balance of C, S, Na, and Mg would only reduce
the C~II$^*$ excitation by 30$\%$ - an amount well within the
uncertainty of N(C~II$^*$).  In other words, the high LLCC hydrogen
density in the HD~85259 sightline makes its C~II fine-structure
excitation relatively insensitive as an indicator of the cloud's
electron density.

\subsection{Molecular Hydrogen in the LLCC}

Unlike the cases of C, S, Na, and Mg, the ionization balance of
Cl in diffuse clouds is complicated by an exothermic reaction between
Cl~II and H$_2$ which can lead to the predominance of the ``trace''
neutral Cl~I if the abundance of H$_2$ is significant
\citep{jur78,moo11}.  Based on the measured Cl~I and Zn~II columns
(Table~2), the \citet{lod03} solar Cl abundance, and the Cl dust
depletion ($+$0.07~dex) appropriate for an $F_*$~$=$~0.3 cloud
\citep{jen09a}, the LLCC Cl~II column density in the HD~85259
sightline is 5.70$\times$10$^{11}$~cm$^{-2}$.  Assuming an H$_2$-free
cloud where the Cl photoionization equilibrium is governed by
$\alpha$(Cl~II)~$=$~9.91$\times$10$^{-11}$~cm$^3$~s$^{-1}$
\citep{ver99} and
$\Gamma$(Cl~I)~$=$~3.59$\times$10$^{-10}$~s$^{-1}$
\citep{dra11,mat83},
the Cl~I and Cl~II column densities toward HD~85259 would indicate
an LLCC electron density of 10.6~cm$^{-3}$.  Although this electron
density leads to a value of $n_e$/$n_H$ ($\approx$0.003) that is
consistent with those of other diffuse clouds \citep{wel03}, it is
$\approx$100 times greater than the mean LLCC electron density
determined from the photoionization balance of C, S, Na, and Mg.
Alternatively, a modest abundance of H$_2$ could explain the
observed LLCC Cl~I column toward HD~85259.  Assuming
$n_e$~$=$~0.1~cm$^{-3}$ and utilizing a rate constant of
k~$=$~7$\times$10$^{-10}$~cm$^3$~s$^{-1}$ for the reaction between
Cl~II and H$_2$ \citep{wel99}, the required H$_2$ density would
be 1.5~cm$^{-3}$.  This H$_2$ density is certainly modest in
comparison to the hydrogen density ($n_H$~$\approx$~4000~cm$^{-3}$)
measured from the C~I fine-structure excitation and would
indicate an LLCC H$_2$ column density of
$\approx$3$\times$10$^{15}$~cm$^{-2}$ in the HD~85259 sightline.

The modest LLCC H$_2$ column density inferred from the Cl
ionization balance toward HD~85259 is consistent with the lack
of ultraviolet CO absorption in this sightline.  Specifically, we
measure a 2$\sigma$ upper limit of 2.5$\times$10$^{12}$~cm$^{-2}$
on the CO column density based on the absence of the
[A-X](4-0)R(0) $\lambda$1419.044 line in our STIS spectra of
HD~85259.  The interstellar CO/H$_2$ column density ratio has
been found to vary from $\approx$10$^{-4}$ for dense clouds to
$\approx$10$^{-5}$ for translucent clouds to $\approx$10$^{-7}$
for diffuse clouds \citep{bur07}.  Even with the assumption of
the highest of these values for the LLCC, the inferred H$_2$
column density toward HD~85259 implies a CO column density
($\approx$3$\times$10$^{11}$~cm$^{-2}$) that is well below our
STIS upper limit.

\subsection{Solar-System-Scale Na~I Structure in the LLCC}

A key implication of the high LLCC gas pressures measured toward
HD~85259 and HD~83023 is that the densest region of the cloud must
be very thin.  In the case of the HD~85259 sightline, a comparison
of the pressure-inferred hydrogen volume density
($n_H$~$\approx$~4000~cm$^{-3}$) and the measured H~I column density
yields an LLCC H~I thickness of $\approx$200~AU\@.  However, it is
important to note that the spatial distribution of trace neutrals
like C~I in diffuse clouds is not likely to match that of H~I.  As
discussed by \citet{lau03}, since trace neutrals are biased by the
recombination rates to form most abundantly near the peaks of the
hydrogen density distribution within a cloud, their abundances will
fluctuate more strongly on smaller length scales than those of H~I
and dominant ions.  Consequently, in the presence of a sharp
density peak, the LLCC gas pressure measured
by the C~I fine-structure excitation toward HD~85259 would be more
representative of the density peak where C~I is most abundant than
that of the full H~I extent of the LLCC.  In other words, the C~I
thickness of the LLCC toward HD~85259 could be somewhat less than
200~AU while the H~I thickness could be appreciably greater.  If
the trace neutral ``layer'' is indeed this thin, one would expect
to observe column density fluctuations in the trace neutrals on
very small transverse scales across the LLCC\@.  Testing for such
solar-system-scale structure is possible through multi-epoch Na~I
absorption-line observations of high proper-motion background stars.

As reviewed by \citet{lau07}, about 15$\%$ of the 40 diffuse cloud
sightlines studied for temporal Na~I absorption variations have
revealed evidence for structure on scales of 50~AU or less.  The
sightlines exhibiting such variations appear to be preferentially
associated with intervening H~I shells.  Variations in dominant
ion absorption on these scales have been observed in only one case
to date~-~the CPD~-59~2603 sightline through the Carina Nebula
\citep{dan01}.  Among the clouds exhibiting solar-system-scale
Na~I structure, those in the sightlines toward HD~32040
\citep{lau00}, $\rho$~Leo \citep{lau03}, and HD~219188
\citep{wel07} have been evaluated through observations of their
C~I fine-structure excitation.  The derived pressures of these
clouds range from 2000~cm$^{-3}$~K to an upper limit of
20,000~cm$^{-3}$~K with inferred hydrogen densities ($n_H$)
between 20~cm$^{-3}$ and 200~cm$^{-3}$.  In the case of the
HD~219188 cloud, \citet{wel07} was able to measure the C~I
excitation multiple times with {\it HST} over a 9 yr time span
and found that $n_H$ increased from 20~cm$^{-3}$ to 45~cm$^{-3}$
as N(C~I) and N(Na~I) increased over a transverse spatial scale
somewhere between 10 and 200~AU (scale uncertainty due to unknown
distance of the intervening cloud).  Based on these results and
the lack of dominant ion differences, observations of temporal
Na~I variations have generally been interpreted as arising
from solar-system-scale cloud fluctuations in $n_H$ and/or $n_e$
rather than in N(H~I).

Among the 16 stars found by \citet{mey06} to exhibit narrow Na~I
absorption corresponding to the LLCC, HD~84937 has by far the
largest proper motion (0.860~arcsec~yr$^{-1}$).  In Figure~14,
we compare the initial 2006~January Na~I spectrum of this star
with that taken in 2009~January.  While the broader stellar Na~I
absorption at $v_{LSR}$~$=$~-21~km~s$^{-1}$ is identical in
both spectra, the LLCC Na~I absorption exhibits a significant
difference in strength between the two epochs.  Applying a
two-component fit to the LLCC Na~I D$_2$ and D$_1$ profiles in
both spectra (similar to the HD~85259, HD~83023, and HD~84722
fits), we find that this difference is entirely associated
with the narrow LLCC component.  Specifically, the Na~I column
density corresponding to this component increases by 40$\%$ from
2006 (1.24$\pm$0.07$\times$10$^{11}$~cm$^{-2}$) to
2009 (1.72$\pm$0.08$\times$10$^{11}$~cm$^{-2}$).  Assuming a
distance of 17.8~pc for the LLCC (midway between the $>$11.3~pc
and $<$24.3~pc constraints of \citet{peek11a}), this three-year
increase in N(Na~I) toward HD~84937 occurred over a transverse
distance of 46~AU across the cloud.  This detection of
solar-system-scale Na~I structure is uniquely accurate in scale
in that it is based on a dominant well-measured stellar proper
motion (the modeled LLCC proper motion \citep{hau12} is much
smaller) and a well-constrained cloud distance (the quoted
scales in previous cases have been mostly upper limits based on
the background star distances).  In contrast, there is no
convincing evidence of temporal Na~I absorption variations in
our multi-epoch observations of HD~85259, HD~83023, and
HD~84722.  However, the much smaller proper motions of these
stars translate to respective transverse distances of only
1.4, 2.2, and 2.0~AU across the LLCC over the course of
our observations.  Also, the greater strength of the
LLCC Na~I lines in these sightlines as compared to those toward
HD~84937 makes their equivalent widths less sensitive to
comparable column density differences.  Nevertheless, the
HD~84937 results provide ample motivation to continue monitoring
the Na~I absorption toward the highest proper motion LLCC
background stars among the \citet{mey06} sample.

The detection of Na~I structure on a scale of 46~AU toward
HD~84937 indicates that the trace neutrals in this
LLCC sightline are abundant in a very thin layer.  It is thus
consistent with the sharp density peak and sheet-like LLCC
geometry implied by the high gas pressures measured toward
HD~85259 and HD~83023.  At the assumed 17.8~pc distance of
the LLCC, the HD~85259 and HD~83023 sightlines are separated by
1.4~pc (4.4$\arcdeg$), the HD~83023 and HD~84937 sightlines are
separated by 1.0~pc (3.2$\arcdeg$), and the HD~85259 and HD~84937
sightlines are separated by 0.6~pc (2.0$\arcdeg$).  Given these
separations, it certainly appears that the cold, highly-pressurized,
low~N(H~I) gas observed toward HD~85259 and HD~84937 is representative
of the LLCC and not a localized fluctuation within the cloud.
Understanding the global overpressure of the LLCC gas
with respect to the Local Bubble environment then
requires either a non-equilibrium solution where the LLCC is a
transient phenomenon or something like the \citet{vaz06}
colliding-cloud model where the LLCC can be sustained indefinitely
as a thin, cold, dense interface between converging flows of
warmer gas.

From a broader perspective, it is interesting that the coldest,
densest, low-column-density interstellar cloud now known in the
Galaxy is located only $\approx$20~pc away from the Sun.  It is
either just a fortunate coincidence to have such an unusual cloud
nearby for close study or a manifestation of a selection effect
that works against detecting such clouds at greater distances.
In support of the former, the characteristics of the LLCC not only
stand out in comparison with samples of many diffuse clouds
(such as the comprehensive C~I study of \citet{jen11}), its
overpressure could be interpreted in terms of a transient
phenomenon.  Yet, \cite{jen11} have found evidence for a weak,
but pervasive contribution from high-pressure
(log($p/k$)~$>$~5.5), high-temperature ($T$~$>$~80~K) gas in
their measured distribution of interstellar pressures.  The weak
admixture of this high-$f_1$, high-$f_2$ gas stands out in their
C~I ($f_1$, $f_2$) diagnostic plot by pulling the net
($f_1$, $f_2$) values of the measured gas parcels above the
theoretical curves for the dominant lower pressure gas.  It is
important to note that a similar weak, pervasive admixture
of high-pressure, low-temperature gas (like that of the LLCC)
would not stand out in the \cite{jen11} C~I diagnostics.  This
high-$f_1$, low-$f_2$ gas would pull the net ($f_1$, $f_2$)
values of the measured gas parcels to the right on slopes similar
to that of the theoretical curves for individual clouds of
lower pressure gas.  In other words, it is possible that
LLCC-type clouds may not be uncommon in the Galaxy.

In support of this possibility, it is vital to recognize
that the radio, optical, and UV study of the nearby LLCC
has been aided by the modest interstellar gas background toward
the Galactic halo - identification of LLCCs at greater distances
in the Galactic disk would be greatly complicated by both foreground
(radio, optical, UV) and background (radio) gas with overlapping
velocity signatures.  Furthermore, despite its large angular
size, the total H~I mass of the LLCC is only $\approx$1~M$_{\sun}$.
A distant, similar-sized LLCC would
have a much less noticeable angular extent.  For example, at 1000~pc,
the LLCC would cover only a 12$\arcmin$~$\times$~2.5$\arcmin$ patch
of sky.  Given the observation of solar-system-scale Na~I structure
toward HD~84937, it is tempting to consider this characteristic as
a potential LLCC-type signature in the other sightlines where it has
been observed.  However, none of these sightlines have yet revealed
C~I pressures as high as those measured in the LLCC despite their
preferential association with H~I shells.  A key step toward a
broader understanding of the LLCC will come with the measurement
of the distances, pressures, and structure of the other cold clouds
comprising the 70$\arcdeg$-long \citet{hau10} H~I ribbon stretching
from Gemini through Leo to Sextans.



\acknowledgments

We are especially grateful to the referee, Edward Jenkins, whose
detailed and constructive review was very helpful in improving the paper.
We would also like to thank Urmas Haud for his calculation of the
LLCC transverse velocity, Enrique Vazquez-Semadeni for his valuable
comments on the manuscript, and Daryl Willmarth for his assistance with
the KPNO observations.  Support for program GO 11736 was provided by NASA
through a grant from the Space Telescope Science Institute, which is
operated by the Association of Universities for Research in Astronomy,
Incorporated, under NASA contract NAS5-26555.  JEGP was supported by
HST-HF-51295.01A, provided by NASA through a Hubble Fellowship grant
from STScI, which is operated by AURA under NASA contract NAS5-26555.



{\it Facilities:} \facility{HST (STIS)}, \facility{KPNO (Coude Feed)}




\clearpage



\begin{figure}
\epsscale{0.7}
\plotone{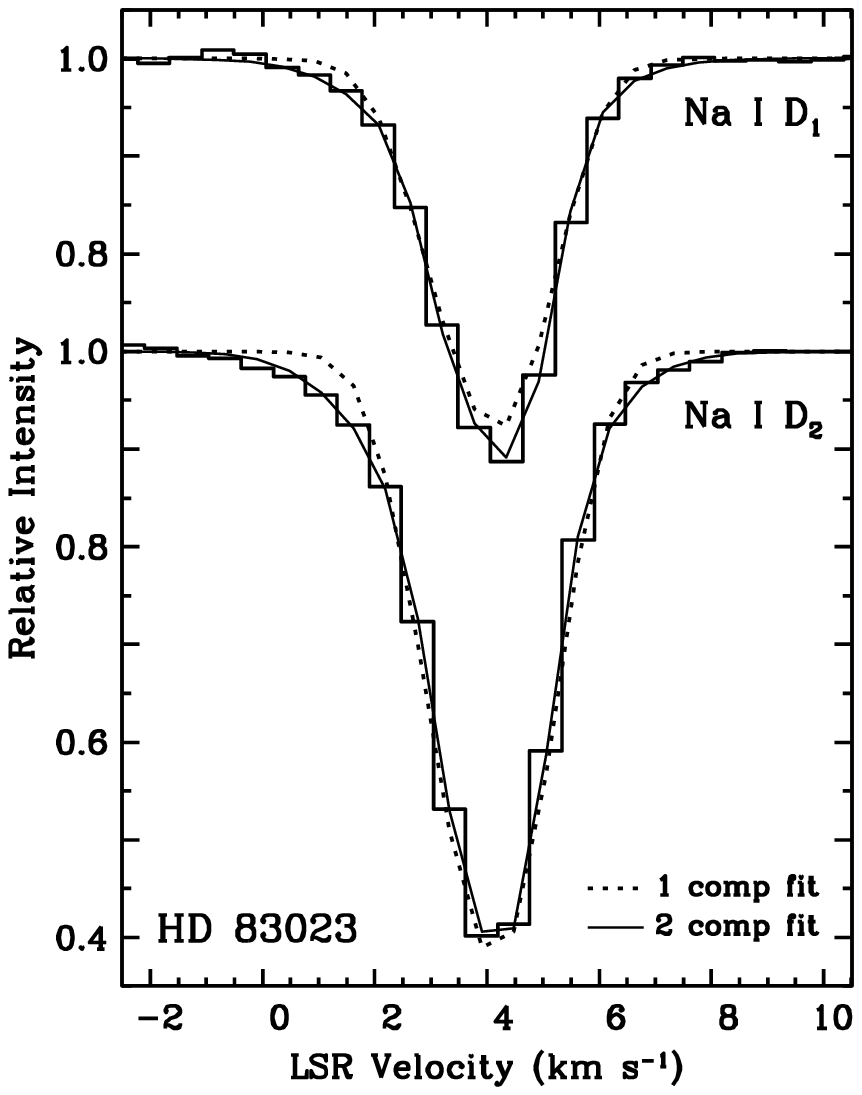}
\caption{High-resolution ($\Delta$v~$=$~1.3~km~s$^{-1}$) KPNO coude
feed spectra of the interstellar Na~I D$_1$ $\lambda$5895.924 and
D$_2$ $\lambda$5889.951 absorption profiles toward HD~83023.  The
dotted lines through the data represent the best single velocity
component Voigt fit to both of the observed profiles (the oscillator
strength of the D$_2$ line is twice that of the D$_1$ line).  The
smooth lines through the data constitute the best two-component
Voigt fit to the observed profiles.  The central LSR velocities,
Na~I column densities, and line widths of the components associated
with both fits are listed in Table~1.
\label{fig1}}
\end{figure}

\clearpage

\begin{figure}
\epsscale{0.7}
\plotone{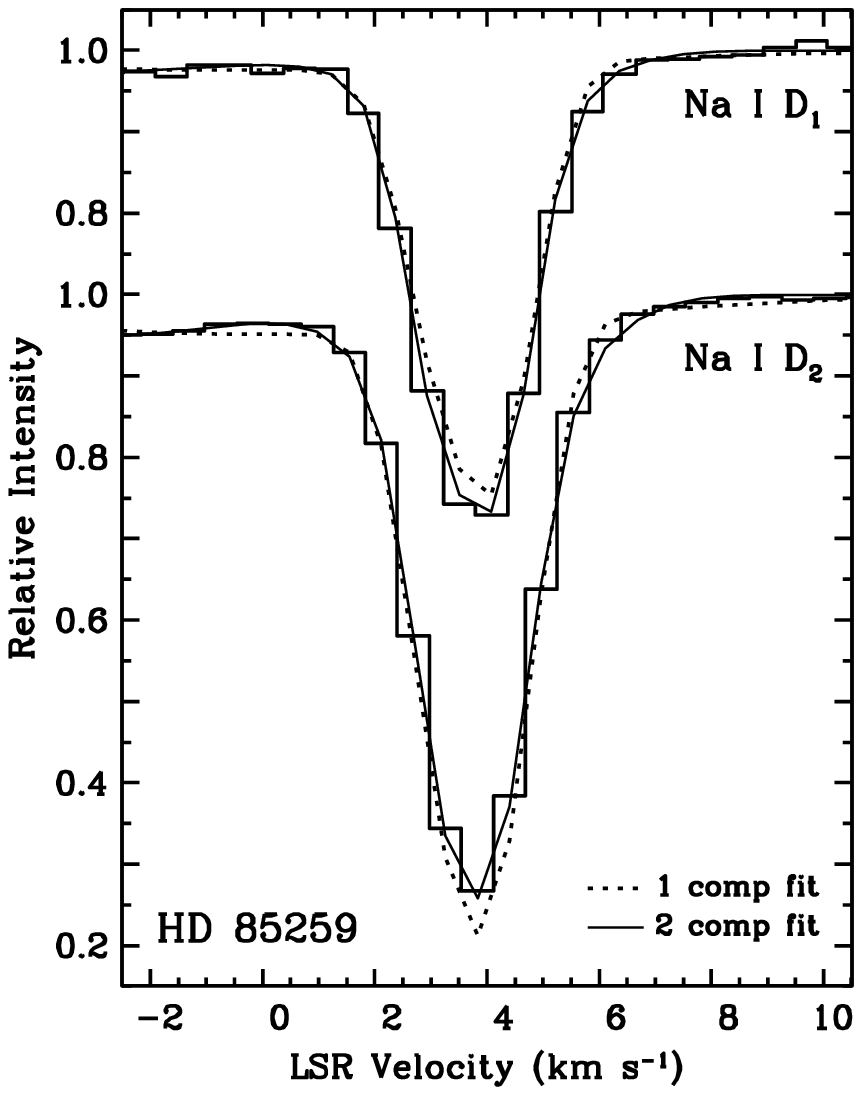}
\caption{High-resolution ($\Delta$v~$=$~1.3~km~s$^{-1}$) KPNO coude
feed spectra of the interstellar Na~I D$_1$ $\lambda$5895.924 and
D$_2$ $\lambda$5889.951 absorption profiles toward HD~85259.  The
dotted lines through the data represent the best single velocity
component Voigt fit to both of the observed profiles (the oscillator
strength of the D$_2$ line is twice that of the D$_1$ line).  The
smooth lines through the data constitute the best two-component
Voigt fit to the observed profiles.  The central LSR velocities,
Na~I column densities, and line widths of the components associated
with both fits are listed in Table~1.
\label{fig2}}
\end{figure}

\clearpage

\begin{figure}
\epsscale{0.7}
\plotone{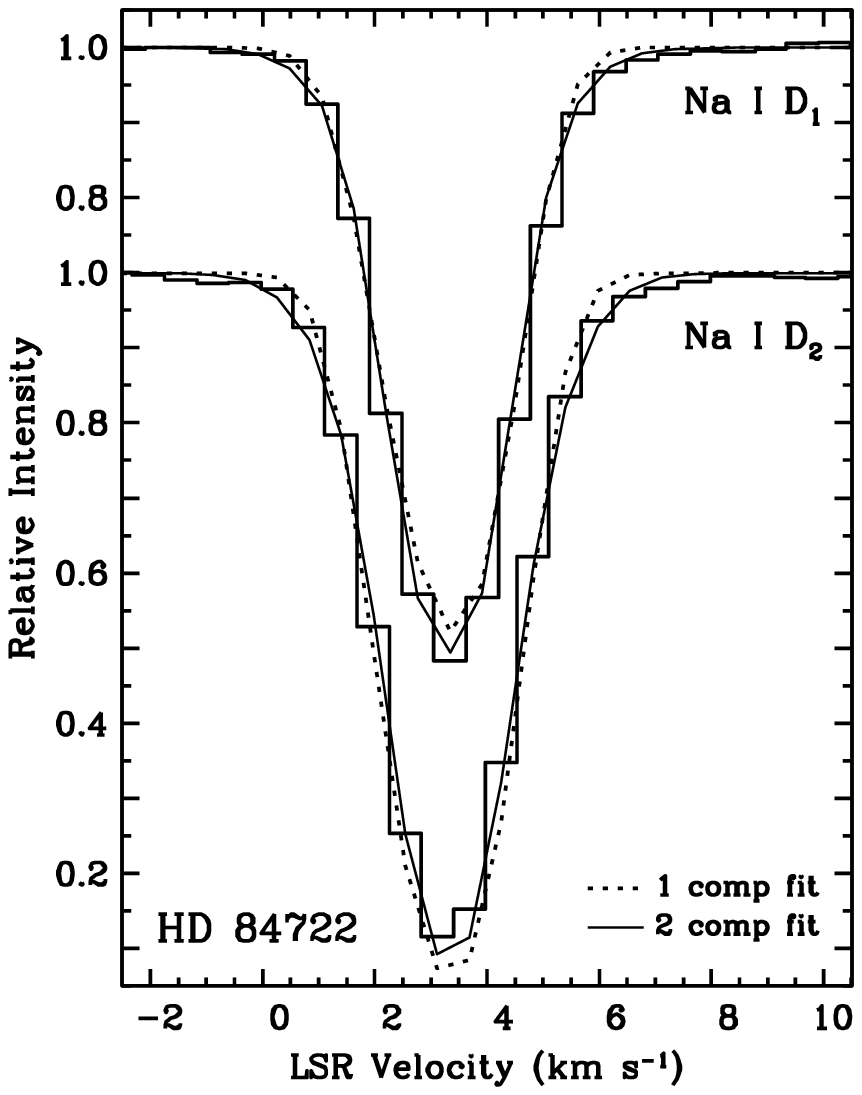}
\caption{High-resolution ($\Delta$v~$=$~1.3~km~s$^{-1}$) KPNO coude
feed spectra of the interstellar Na~I D$_1$ $\lambda$5895.924 and
D$_2$ $\lambda$5889.951 absorption profiles toward HD~84722.  The
dotted lines through the data represent the best single velocity
component Voigt fit to both of the observed profiles (the oscillator
strength of the D$_2$ line is twice that of the D$_1$ line).  The
smooth lines through the data constitute the best two-component
Voigt fit to the observed profiles.  The central LSR velocities,
Na~I column densities, and line widths of the components associated
with both fits are listed in Table~1.
\label{fig3}}
\end{figure}

\clearpage

\begin{figure}
\epsscale{0.7}
\plotone{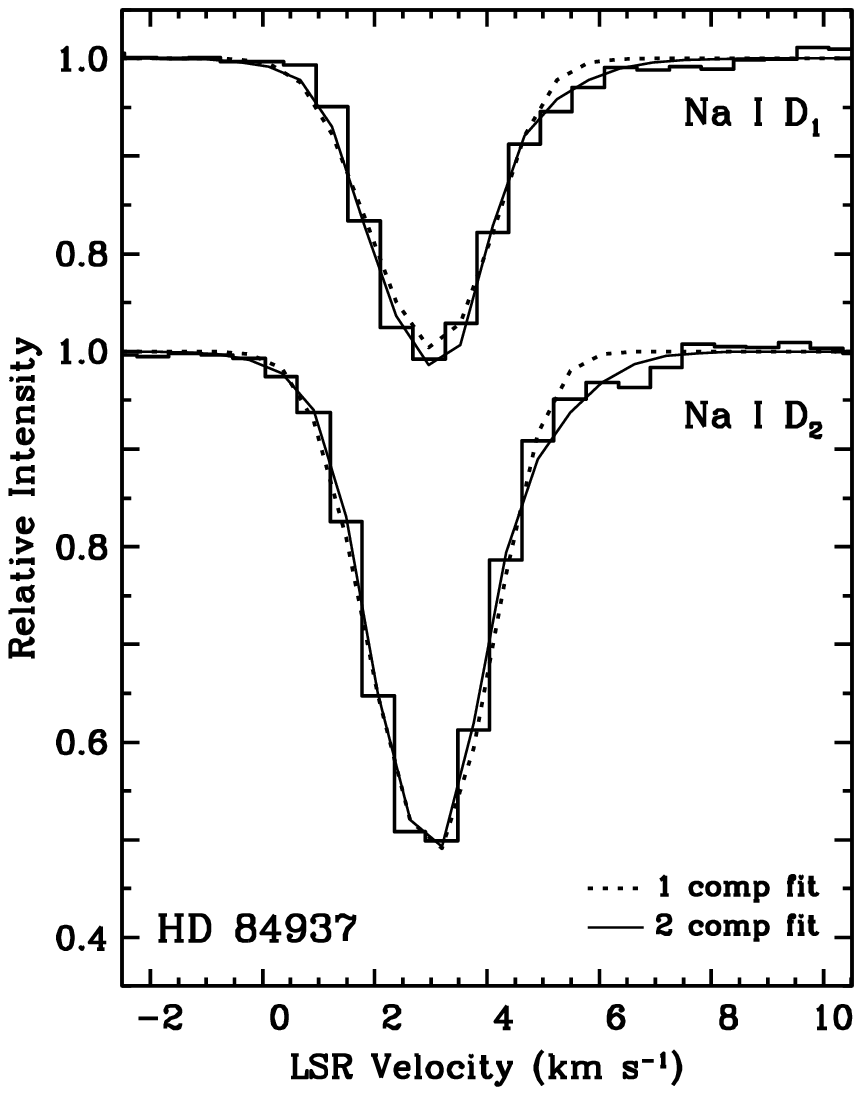}
\caption{High-resolution ($\Delta$v~$=$~1.3~km~s$^{-1}$) KPNO coude
feed spectra of the interstellar Na~I D$_1$ $\lambda$5895.924 and
D$_2$ $\lambda$5889.951 absorption profiles toward HD~84937.  The
dotted lines through the data represent the best single velocity
component Voigt fit to both of the observed profiles (the oscillator
strength of the D$_2$ line is twice that of the D$_1$ line).  The
smooth lines through the data constitute the best two-component
Voigt fit to the observed profiles.  The central LSR velocities,
Na~I column densities, and line widths of the components associated
with both fits are listed in Table~1.
\label{fig4}}
\end{figure}

\clearpage

\begin{figure}
\epsscale{1.0}
\plotone{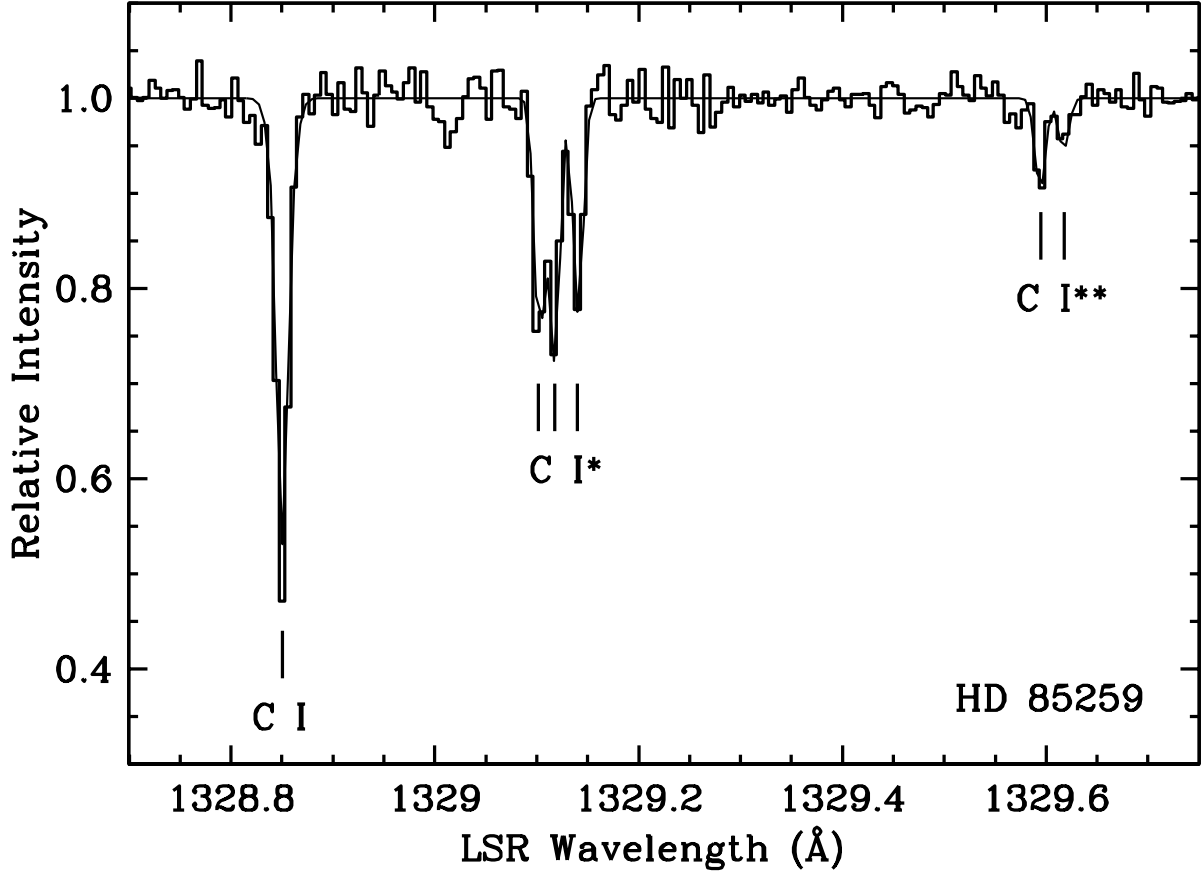}
\caption{High-resolution ($\Delta$v~$=$~2.75~km~s$^{-1}$) {\it HST}
STIS echelle spectra of the interstellar C~I 1329~\AA\ multiplet toward
HD~85259.  The spectra reveal LLCC C~I absorption arising from the
ground fine-structure state (C~I $\lambda$1328.833), the first
excited state (C~I$^*$ $\lambda\lambda$1329.085, 1329.100,
1329.123), and the second excited state (C~I$^{**}$
$\lambda\lambda$1329.578, 1329.601).  The smooth line through the
data represents the best simultaneous fit in terms of central LSR
velocity, column density, and line width to all of the C~I, C~I$^*$,
and C~I$^{**}$ lines observed in the STIS spectra of HD~85259.
These C~I, C~I$^*$, and C~I$^{**}$ column densities are listed
in Table~2.
\label{fig5}}
\end{figure}

\clearpage

\begin{figure}
\epsscale{1.0}
\plotone{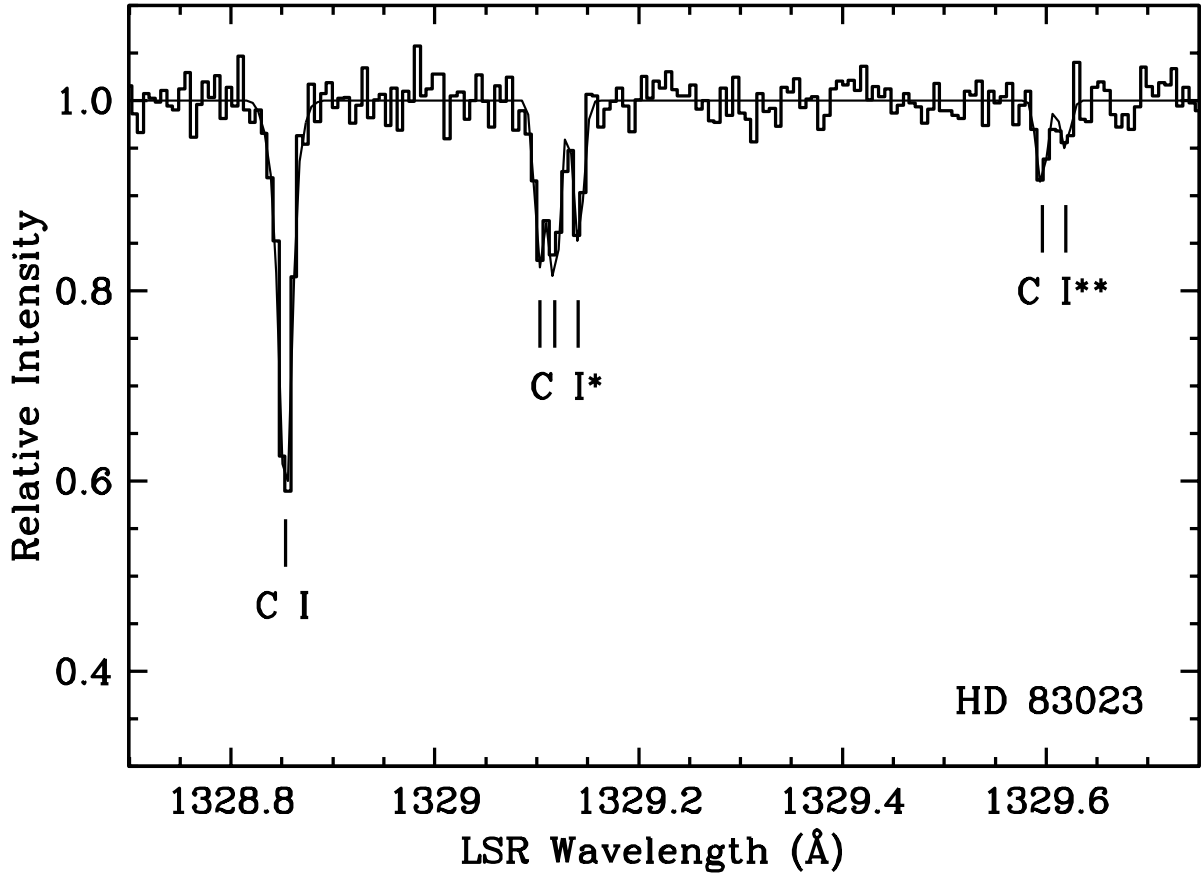}
\caption{Same as for Fig. 5, except for the star HD~83023.
\label{fig6}}
\end{figure}

\clearpage

\begin{figure}
\epsscale{0.6}
\plotone{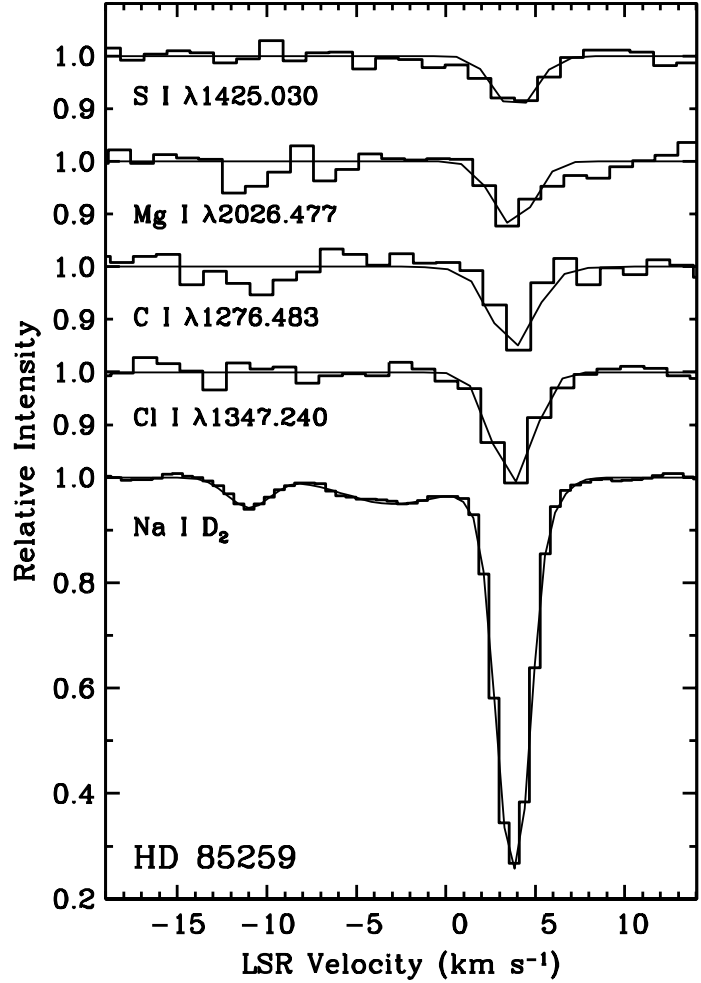}
\caption{Comparison of the LLCC Na~I D$_2$ absorption profile with
those of other trace neutrals observed in the STIS spectra of
HD~85259.  The Na~I absorption blueward of the strong LLCC feature
at v$_{LSR}$~$=$~3.9~km~s$^{-1}$ arises from the neutral gas
boundary of the Local Bubble at a distance between 100 and 150~pc.
The C~I $\lambda$1276.483 profile represents the weakest
LLCC C~I line detectable in the HD~85259 STIS spectra.  Its
oscillator strength is 7.6 times smaller than that of the
C~I $\lambda$1328.833 line shown in Figure~5.  The illustrated
fits to the S~I, Mg~I, C~I, Cl~I, and Na~I profiles reflect
the LLCC column densities listed in Table~2 for the HD~85259
sightline.
\label{fig7}}
\end{figure}

\clearpage

\begin{figure}
\epsscale{0.6}
\plotone{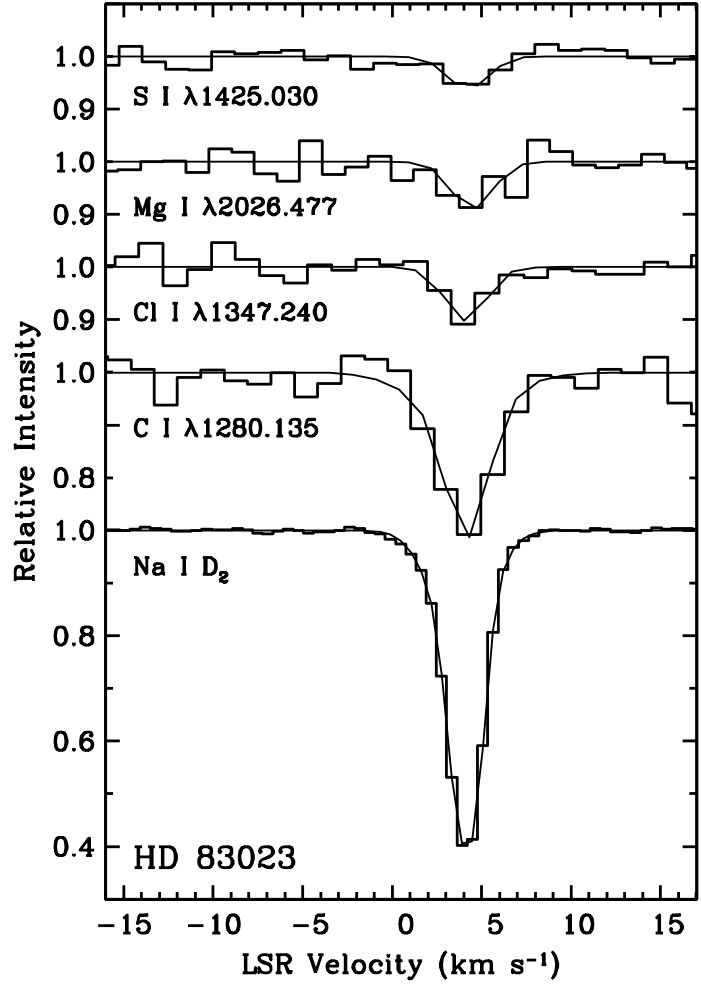}
\caption{Comparison of the LLCC Na~I D$_2$ absorption profile with
those of other trace neutrals observed in the STIS spectra of
HD~83023.  The C~I $\lambda$1280.135 profile represents the weakest
LLCC C~I line detectable in the HD~83023 STIS spectra.  Its 
oscillator strength is 1.9 times smaller than that of the
C~I $\lambda$1328.833 line shown in Figure~6.  The illustrated
fits to the S~I, Mg~I, Cl~I, C~I, and Na~I profiles reflect
the LLCC column densities listed in Table~2 for the HD~83023
sightline.  
\label{fig8}}
\end{figure}

\clearpage

\begin{figure}
\epsscale{0.4}
\plotone{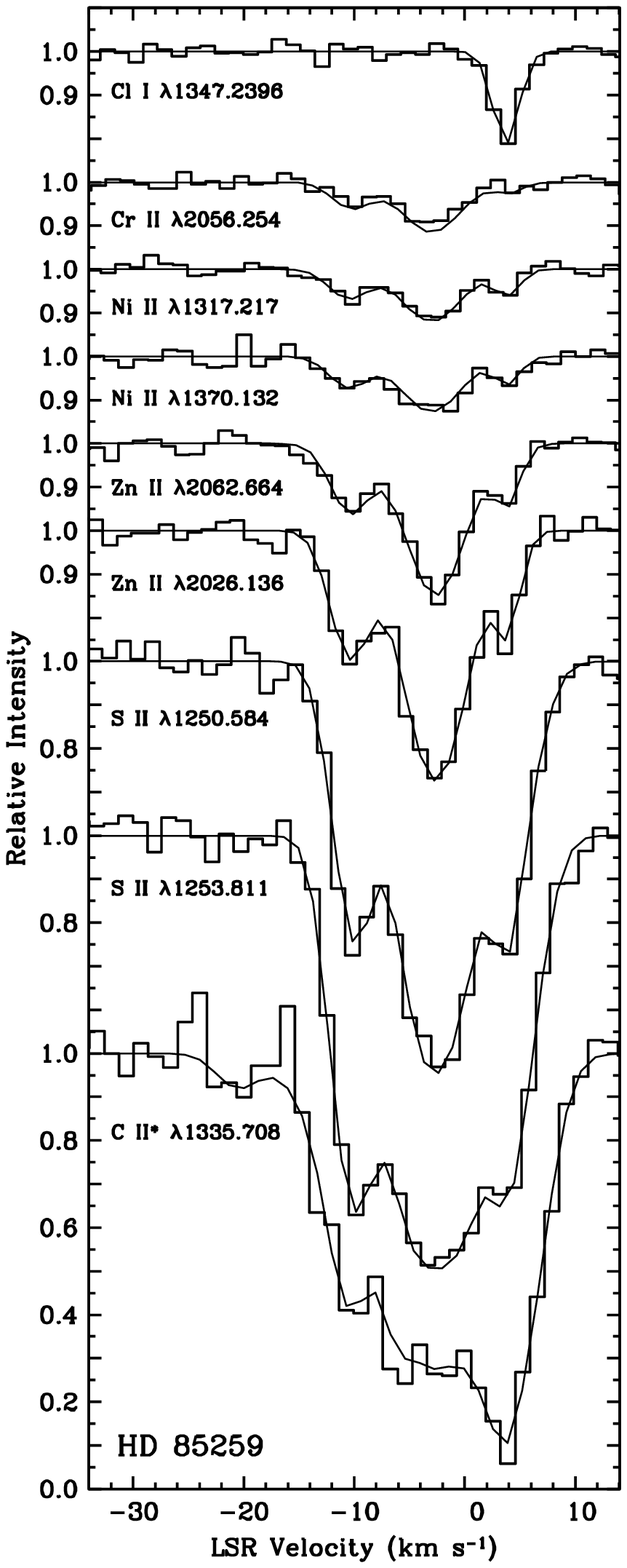}
\caption{Comparison of the Cl~I $\lambda$1347.2396 absorption
profile with those of selected dominant ions in the STIS spectra
of HD~85259.  Unlike the Cl~I profile where the LLCC absorption
stands alone, the Cr~II, Ni~II, Zn~II, S~II, and C~II$^*$
profiles all show significant blueward absorption arising from
gas beyond the LLCC.  The illustrated fits to these profiles
reflect the LLCC column densities listed in Table~2 for the
HD~85259 sightline.
\label{fig9}}
\end{figure}

\clearpage

\begin{figure}
\epsscale{0.4}
\plotone{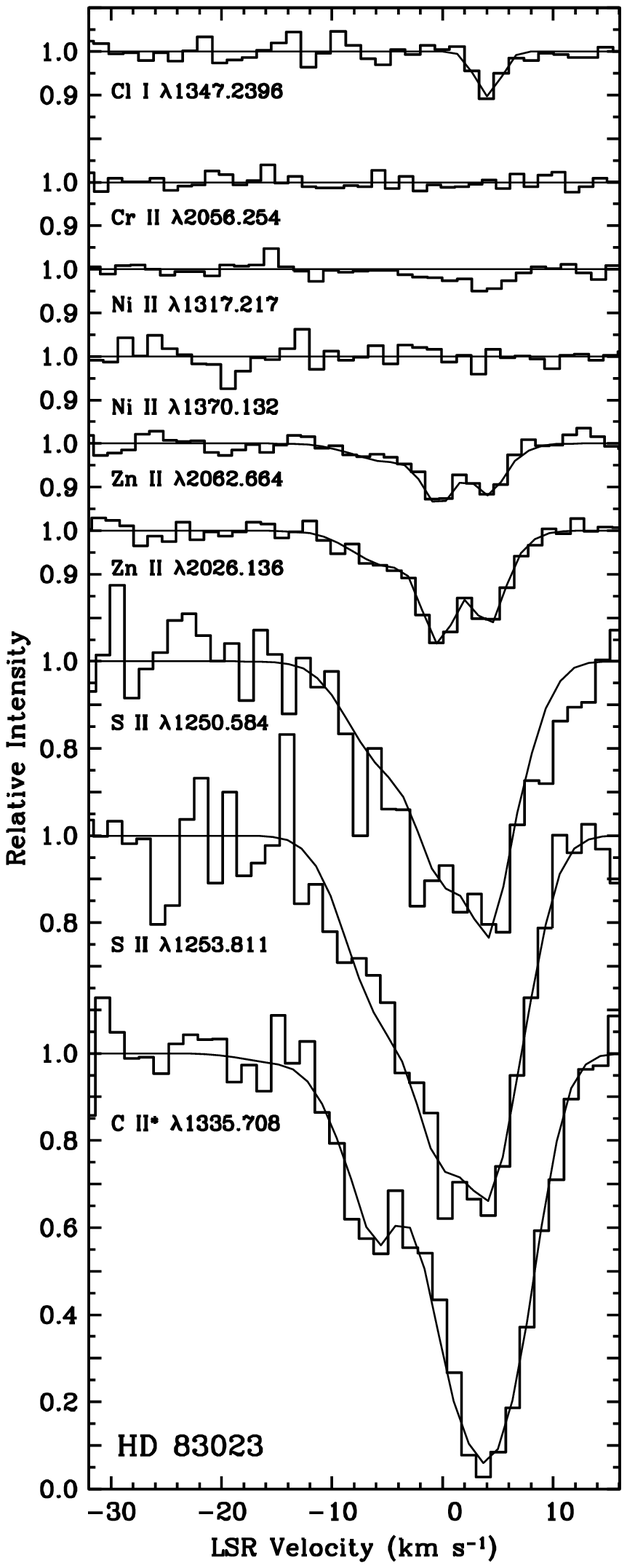}
\caption{Comparison of the Cl~I $\lambda$1347.2396 absorption
profile with those of selected dominant ions in the STIS spectra
of HD~83023.  Unlike the Cl~I profile where the LLCC absorption
stands alone, the Zn~II, S~II, and C~II$^*$
profiles all show significant blueward absorption arising from
gas beyond the LLCC.  The illustrated fits to these profiles
reflect the LLCC column densities listed in Table~2 for the
HD~83023 sightline.
\label{fig10}}
\end{figure}

\clearpage

\begin{figure}
\epsscale{0.6}
\plotone{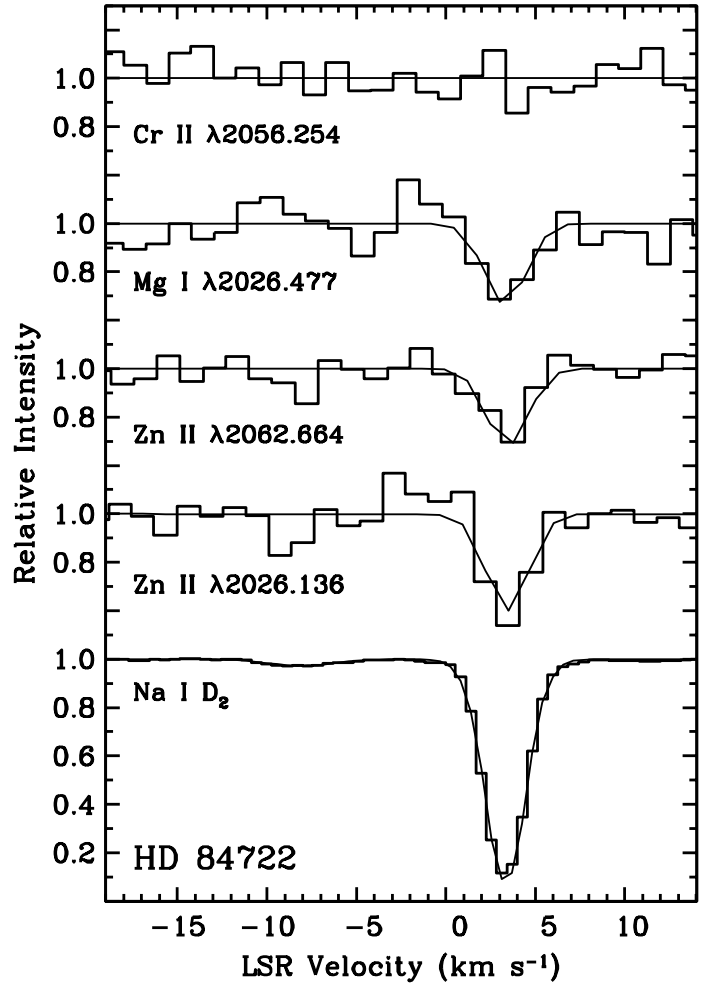}
\caption{Comparison of the LLCC Na~I D$_2$ absorption profile
with those of Cr~II, Mg~I, and Zn~II in the STIS spectra of
HD~84722.  The illustrated fits to the Mg~I, Zn~II, and Na~I
profiles reflect the LLCC column densities listed in Table~2
for the HD~84722 sightline.
\label{fig11}}
\end{figure}

\clearpage

\begin{figure}
\epsscale{1.0}
\plotone{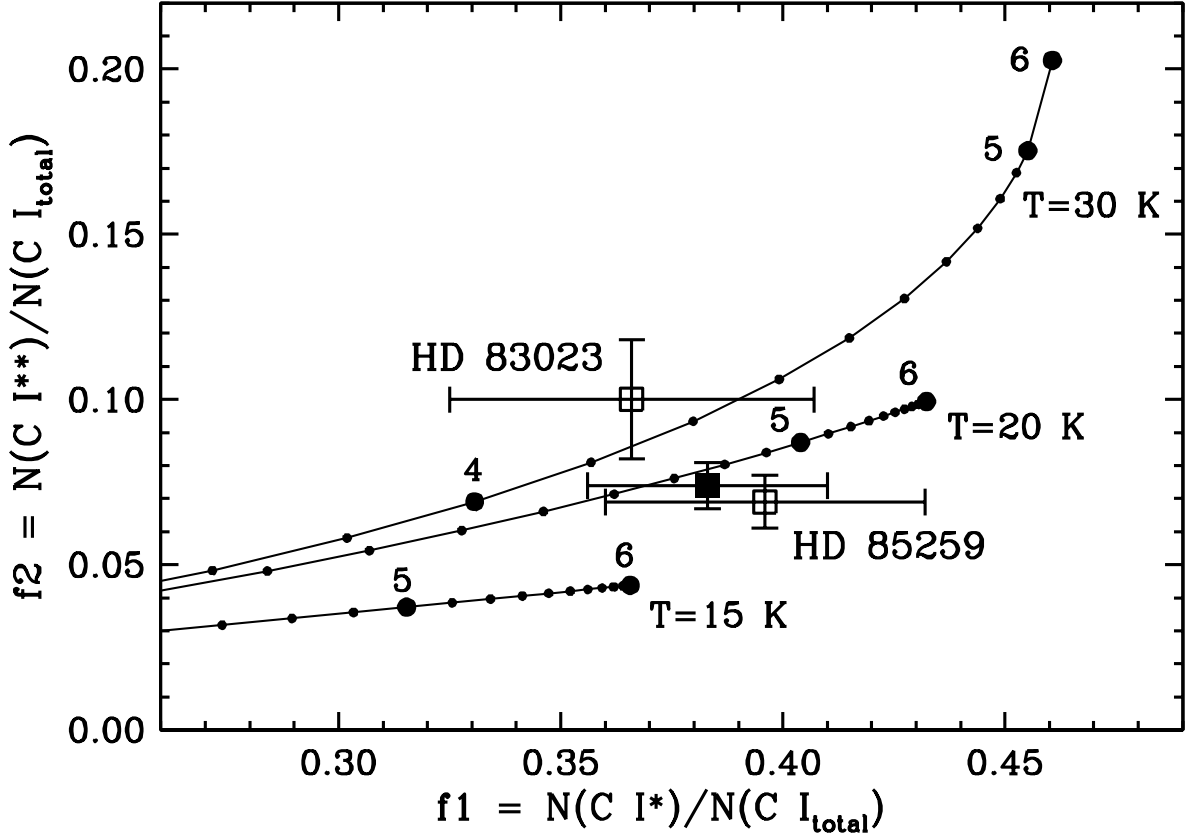}
\caption{The LLCC C~I fine-structure excitation toward HD~85259
and HD~83023 in relation to that expected in gas clouds of various
pressures at temperatures of 15, 20, and 30~K.  The three curves
trace the derived ($f1$,~$f2$) values for each of these temperatures
with the pressure points on the curves denoting 0.1~dex
increments in log (P/k) between the labeled values.  Given that
the LLCC has a temperature of $\approx$20~K, these curves clearly
indicate that the LLCC gas pressure toward both HD~85259 and
HD~83023 is well in excess of 10,000~cm$^{-3}$~K.  Their
weighted mean ($f1$,~$f2$) value (denoted by the black square)
falls almost exactly on the
20~K curve at a pressure of 60,000~cm$^{-3}$~K\@.
\label{fig12}}
\end{figure}

\clearpage

\begin{figure}
\epsscale{0.5}
\plotone{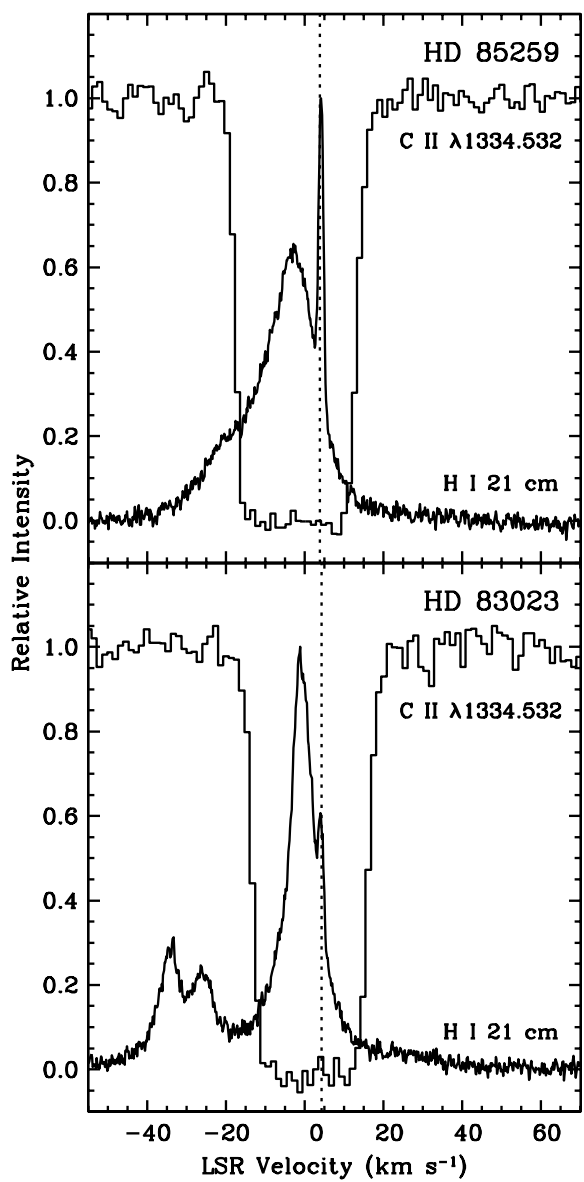}
\caption{Comparison of the C~II $\lambda$1334.532 absorption and
H~I 21~cm emission in the HD~85259 and HD~83023 sightlines.  The
GALFA H~I spectra have spatial and velocity resolutions of
4$\arcmin$ and 0.184~km~s$^{-1}$.  The LSR velocity of the LLCC
Na~I absorption toward HD~85259 and HD~83023 is denoted with a
dotted line through the displayed spectra.  The absence of any
C~II absorption lines outside the single saturated feature in
these spectra indicates that there are no foreground warm cloud
pairs in the LLCC vicinity with H column densities greater
than 10$^{16}$~cm$^{-2}$ and a radial velocity differential greater
than 35~km~s$^{-1}$.
\label{fig13}}
\end{figure}

\clearpage

\begin{figure}
\epsscale{0.7}
\plotone{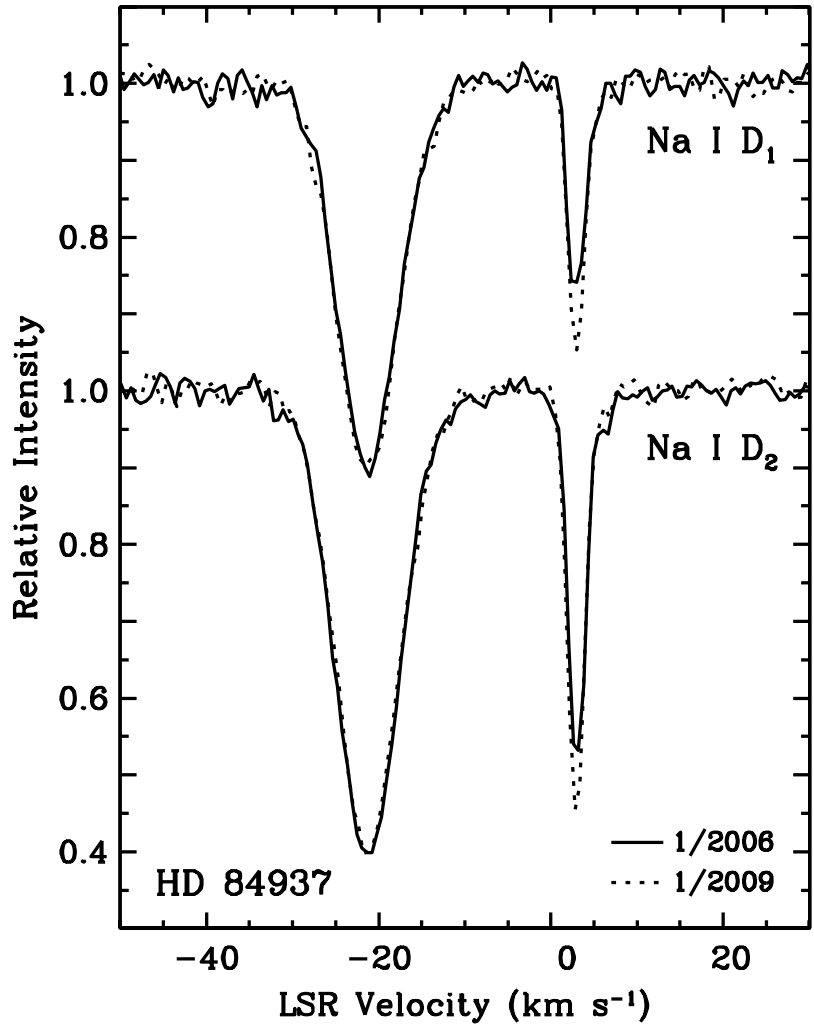}
\caption{Comparison of the Na~I~D absorption profiles toward
HD~84937 in KPNO Coude Feed spectra taken in 2006 January and
in 2009 January.  While the broader stellar Na~I absorption at
$v_{LSR}$~$=$~-21~km~s$^{-1}$ is identical in both spectra,
the narrow LLCC Na~I absorption exhibits a significant difference
in strength between the two epochs.  This difference corresponds
to an 40$\%$ increase in the LLCC Na~I column density over the
course of 3 years in the HD~84937 sightline.  Given the
0.860~arcsec~yr$^{-1}$ proper motion of HD~84937 and a distance
estimate of 17.8~pc for the LLCC, this increase occurred over
a transverse distance of 46~AU across the cloud.
\label{fig14}}
\end{figure}

\clearpage





\begin{deluxetable}{cccc}
\tabletypesize{\scriptsize}
\tablecolumns{4}
\tablecaption{THE LLCC Na~I AND H~I PROFILE FITS\label{tbl-1}}
\tablewidth{0pt}
\tablehead{
\colhead{Star} & \colhead{Column Density\tablenotemark{a}} &
\colhead{$b$-value\tablenotemark{b}} & \colhead{LSR Velocity} \\
\colhead{} & \colhead{(cm$^{-2}$)} &
\colhead{(km s$^{-1}$)} & \colhead{(km s$^{-1}$)}}
\startdata
\cutinhead{Single-Component Na I Fit}
HD 83023 & $2.39\times10^{11}$ & 0.84 & $+4.10$ \\
HD 85259 & $3.73\times10^{11}$ & 0.46 & $+3.85$ \\
HD 84722 & $7.55\times10^{11}$ & 0.71 & $+3.38$ \\
HD 84937 & $1.77\times10^{11}$ & 0.86 & $+3.01$ \\
\cutinhead{Double-Component Na I Fit}
HD 83023 & $2.23\times10^{11}$ & 0.24 & $+4.23$ \\
         & $8.99\times10^{10}$ & 1.83 & $+3.87$ \\
HD 85259 & $5.69\times10^{11}$ & 0.24 & $+3.87$ \\
         & $7.54\times10^{10}$ & 1.68 & $+3.93$ \\
HD 84722 & $2.12\times10^{12}$ & 0.24 & $+3.44$ \\
         & $2.29\times10^{11}$ & 1.35 & $+3.37$ \\
HD 84937 & $1.43\times10^{11}$ & 0.24 & $+2.95$ \\
         & $6.95\times10^{10}$ & 1.70 & $+3.37$ \\
\cutinhead{Double-Component H I Fit}
HD 83023 & $3.15\times10^{18}$ & 0.46 & $+4.49$ \\
         & $1.84\times10^{18}$ & 0.37 & $+3.86$ \\
HD 85259 & $1.03\times10^{19}$ & 0.64 & $+3.86$ \\
         & $5.80\times10^{18}$ & 0.54 & $+4.53$ \\
HD 84722 & $2.92\times10^{19}$ & 0.67 & $+3.31$ \\
         & $1.37\times10^{19}$ & 0.49 & $+4.09$ \\
\enddata
\tablenotetext{a}{The measured Na~I column densities are based on
single- and double-component Voigt profile fits to the
1.3~km~s$^{-1}$ resolution optical spectra.  The
measured H~I column densities are based on double-component
Gaussian profile fits to the 0.184~km~s$^{-1}$ resolution
21~cm spectra \citep{peek11a}.}
\tablenotetext{b}{The measured $b$-values represent the Gaussian
line widths of the profile components ($b$~$=$~FWHM/1.665).}
\end{deluxetable}
\clearpage

\begin{deluxetable}{lccc}
\tabletypesize{\scriptsize}
\tablecolumns{4}
\tablecaption{THE LLCC COLUMN DENSITIES\label{tbl-2}}
\tablewidth{0pt}
\tablehead{
\colhead{Species} & \colhead{HD 85259} &
\colhead{HD 83023} & \colhead{HD 84722}}
\startdata
\cutinhead{Narrow Component\tablenotemark{a}}
C I & $1.27\pm0.17\times10^{13}$ & $7.89\pm1.30\times10^{12}$ & \nodata \\
C I$^*$ & $9.42\pm0.84\times10^{12}$ & $5.40\pm0.58\times10^{12}$ & \nodata \\
C I$^{**}$ & $1.65\pm0.13\times10^{12}$ & $1.47\pm0.24\times10^{12}$ & \nodata \\
Na I & $5.69\pm0.17\times10^{11}$ & $2.23\pm0.05\times10^{11}$ & $2.12\pm0.17\times10^{12}$ \\
Mg I & $7.09\pm1.21\times10^{11}$ & $4.99\pm1.33\times10^{11}$ & $3.87\pm1.19\times10^{12}$ \\
S I & $7.37\pm0.93\times10^{11}$ & $4.21\pm1.05\times10^{11}$ & \nodata \\
Cl I & $1.67\pm0.12\times10^{12}$ & $6.44\pm0.92\times10^{11}$ & \nodata \\
\\
C II$^*$ & $\sim3\times10^{13}$ & $\sim1\times10^{13}$ & \nodata \\
Cr II & $1.25\pm0.57\times10^{11}$ & $<1.3\times10^{11}$ & $<5.1\times10^{11}$ \\
Ni II & $9.22\pm1.52\times10^{11}$ & $<8.0\times10^{11}$ & \nodata \\
Zn II & $3.55\pm0.37\times10^{11}$ & $1.41\pm0.42\times10^{11}$ & $1.47\pm0.44\times10^{12}$ \\
\cutinhead{Broad Component\tablenotemark{a}}
C I & $2.38\pm0.30\times10^{12}$ & $3.16\pm0.50\times10^{12}$ & \nodata \\
Na I & $7.54\pm0.36\times10^{10}$ & $8.99\pm0.34\times10^{10}$ & $2.29\pm0.13\times10^{11}$ \\
\\
C II$^*$ & $2.57\pm0.26\times10^{13}$ & $4.66\pm0.41\times10^{13}$ & \nodata \\
S II\tablenotemark{b} & $1.94\pm0.20\times10^{14}$ & $2.82\pm0.28\times10^{14}$ & \nodata \\
Zn II & \nodata & $2.79\pm0.48\times10^{11}$ & \nodata \\
\enddata
\tablenotetext{a}{The column densities are listed in units of cm$^{-2}$.
The quoted errors in the column densities are $\pm1\sigma$ values.
All upper limits are 2$\sigma$ values.}
\tablenotetext{b}{The broad-component S~II column densities toward
HD~85259 and HD~83023 were fit assuming respective narrow-component
S~II column densities of $1.62\times10^{14}$ and
$6.44\times10^{13}$~cm$^{-2}$ (based on the measured narrow-component
Zn II columns, the solar S/Zn abundance ratio, and a Zn depletion into
dust of 0.1 dex relative to S).}
\end{deluxetable}
\clearpage


\end{document}